 \documentstyle[multicol,prc,aps,eqsecnum,epsf]{revtex}
\begin{document}
\draft
%\preprint{\vbox{. \hfill ADP-98-21-T297}}

\title{Weak capture of protons by protons}
\author{R.\ Schiavilla}
\address{Jefferson Lab, Newport News, Virginia 23606 \\
         and Department of Physics, Old Dominion University,
         Norfolk, Virginia 23529}
\author{V.G.J.\ Stoks}
\address{Centre for the Subatomic Structure of Matter,
         University of Adelaide, Adelaide, Australia 5005 \\
         and Physics Division, Argonne National Laboratory,
         Argonne, Illinois 60439}
\author{W.\ Gl\"ockle, H.\ Kamada,\footnote{
               Also Institut f\"ur Kernphysik, Technische Hochschule
               Darmstadt, D-64289 Darmstadt, Germany.}
        and A.\ Nogga}
\address{Institut f\"ur Theoretische Physik II, Ruhr-Universit\"at Bochum,
         D-44780 Bochum, Germany}
\author{J.\ Carlson}
\address{Theoretical Division, Los Alamos National Laboratory,
         Los Alamos, New Mexico 87545}
\author{R.\ Machleidt}
\address{Department of Physics, University of Idaho,
         Moscow, Idaho 83483}
\author{V.R.\ Pandharipande}
\address{Department of Physics, University of Illinois,
         Urbana, Illinois 61801}
\author{R.B.\ Wiringa}
\address{Physics Division, Argonne National Laboratory,
         Argonne, Illinois 60439}
\author{A.\ Kievsky, S.\ Rosati,\footnote{Also Department of Physics,
        University of Pisa, I-56100 Pisa, Italy} and M.\ Viviani}
\address{INFN, Sezione di Pisa, I-56100 Pisa, Italy \\ \phantom{A}}
%\author{S.\ Rosati}
%\address{Department of Physics, University of Pisa, I-56100 Pisa, Italy\\
%         and INFN, Sezione di Pisa, I-56100 Pisa, Italy}
%\author{M.\ Viviani}
%\address{INFN, Sezione di Pisa, I-56100 Pisa, Italy}
\date{published in Phys.\ Rev.\ C {\bf 58}, 1263 (1998).}

\maketitle

\begin{abstract}
The cross section for the proton weak capture reaction
$^1$H($p$,$e^+\nu_e$)$^2$H is calculated with wave functions obtained
from a number of modern, realistic high-precision interactions.
To minimize the uncertainty in the axial two-body current operator, its
matrix element has been adjusted to reproduce the measured Gamow-Teller
matrix element of tritium $\beta$ decay in model calculations using
trinucleon wave functions from these interactions. A thorough analysis
of the ambiguities that this procedure introduces in evaluating the
two-body current contribution to the $pp$ capture is given. Its inherent
model dependence is in fact found to be very weak. The overlap integral
$\Lambda^2(E=0)$ for the $pp$ capture is predicted to be in the range
7.05--7.06, including the axial two-body current contribution, for all
interactions considered.
\end{abstract}
\pacs{21.30.-x, 21.45.+v, 25.10.+s, 95.30.Cq}

\begin{multicols}{2}

\section{INTRODUCTION}
The proton weak capture on protons is the most fundamental process
in stellar nucleosynthesis: it is the first reaction in the $pp$ chain
converting hydrogen into helium, and the principal source for the
production of energy and neutrinos in main-sequence stars.
The theoretical description of this hydrogen-burning reaction, whose
cross section cannot be measured in terrestrial laboratories, was first
given by Bethe and Critchfield~\cite{Bet38}, who showed that the
associated rate was large enough to account for the energy released
by the Sun. Since then, a series of calculations has refined their
original estimate by either computing the required wave functions more
accurately~\cite{Sal52,Bah69,Gou90,Kam94} or by using more realistic
models for the nuclear transition operator~\cite{Gar72,Dau76,Car91}.
We here contribute to this effort by providing an integrated study of
these two aspects with emphasis on a reliable estimate of their
associated theoretical uncertainties.

This paper is divided into seven sections and an appendix.
In Sec.~II we set up the framework for the present study, by providing
expressions for the $pp$ fusion cross section and the required matrix
elements and by summarizing the current ``best'' values for the various
coupling constants, Fermi function, etc. In Secs.~III and IV we give a
fairly detailed description of, respectively, the $pp$ and deuteron
wave functions, as obtained from modern (high-precision) interactions.
The latter include, along with the short-range nuclear part, a complete
treatment of electromagnetic effects up to order $\alpha^2$, $\alpha$
being the fine-structure constant, and accurately reproduce the
measured low-energy $pp$ scattering parameters and deuteron properties.
Sections~V and VI deal with the calculation of the $pp$ cross section
in the approximations, respectively, in which only the one-body or
both the one- and two-body parts of the axial current operator are
retained. In Sec.~VI we also review the evidence, as obtained from an
analysis of tritium $\beta$ decay, for the axial two-body components
(explicit expressions for them are listed in the Appendix). Because of
their model dependence, we adopt the phenomenological approach of
adjusting the cutoff masses in the meson-nucleon vertices and $N$
to $\Delta$ axial coupling constant so as to obtain agreement with
the experimental value for the Gamow-Teller matrix element in tritium
$\beta$ decay. The question of how this procedure impacts the $pp$
cross section is also examined. Finally, in Sec.~VII we summarize
our conclusions, and provide our ``best'' value for the $pp$ overlap
integral at zero energy.

\section{CROSS SECTION}
The spin-averaged total cross section for the $^1$H($p$,$e^+ \nu_e$)$^2$H
reaction can be written in the form~\cite{Car91}
\begin{equation}
\sigma(E) = \frac{1}{(2\pi)^3} \frac{G_V^2} {v_{\rm rel,n.r.}}
       m_e^5 f(E) \sum_{M} | \langle d,M |
     {\bf A}_{-} | pp \rangle|^2. \label{aj}
\end{equation}
Here $G_V$ is the vector coupling constant for which the value
$G_V=(1.149\,39\pm 0.000\,65) \times 10^{-5}$ GeV$^{-2}$ is used, as
obtained from a recent analysis of $ft$ values for superallowed
$0^+\rightarrow 0^+$ transitions~\cite{Har90}; $m_e$ is the electron
mass and $v_{\rm rel,n.r.}$ is the $pp$ relative velocity. The process
is induced by the axial-vector part of the weak interaction Hamiltonian,
and consequently only even parity $pp$ states contribute to the matrix
element.

The naive expression for the Fermi function $f(E)$ is given by 
\begin{eqnarray}
   f(E) &\equiv& \frac{1}{m_e^5} \int\!
       \delta(E+\Delta m-E_\nu-E_e) p_e E_e E_\nu^2 dE_e dE_\nu
                                              \nonumber\\
   &=& \int_1^{(E+\Delta m)/m_e}\! dx\, x
       \sqrt{x^2-1} \left(\frac{E+\Delta m}{m_e}-x\right)^2, \nonumber\\
   &&                                         \label{phsp}
\end{eqnarray}
where $\Delta m=2m_p-m_d=0.931\,25$ MeV~\cite{comm.} ($m_p$ and $m_d$
are the proton and deuteron masses, respectively), $E=k^2/m_p$ is the
c.m.\ incident energy, and the energy of the recoiling deuteron is
neglected. A more refined treatment of the phase-space factor includes the
effect of Coulomb focusing of the emitted $e^+$ wave function~\cite{Bah69}
as well as radiative corrections to the cross section. The latter have
not actually been calculated for the present reaction but have been
estimated to be comparable to those obtained for neutron
decay~\cite{Ade98}. As a result, the Fermi function is parametrized as
\begin{equation}
     f(E)=0.144(1+9.04E),                  \label{feb}
\end{equation}
with $E$ expressed in MeV. At $E=0$ the expression in Eq.~(\ref{phsp})
gives 0.148, which is about 3\% larger than the more accurate estimate
from Eq.~(\ref{feb}).

The deuteron and even parity $pp$ wave functions are written as 
\begin{equation}
  \Psi_d^M({\bf r}) = \left[
  \frac{u(r)}{r} {\cal Y}^{1M}_{01}({\bf \hat{r}}) +
  \frac{w(r)}{r} {\cal Y}^{1M}_{21}({\bf \hat{r}}) \right]
  \zeta_0^0,                                     \label{DW}
\end{equation}
\begin{eqnarray}
  \psi_{\bf k}^{(+)} ({\bf r}) = && 4\pi\sqrt{2}
      \sum_{L\> {\rm even}} \sum_{M_L} {\rm i}^L
      Y^*_{LM_L}({\bf \hat{k}})\,{e^{{\rm i}\delta_L}\over k r}
      \chi_L(r;k)                \nonumber\\
  && \times\ Y_{LM_L}({\bf \hat{r}})\eta_0^0\zeta_1^1, \label{scatt}
\end{eqnarray}
where ${\cal Y}^{JM}_{LS}({\bf \hat{r}})$ are the normalized
eigenfunctions of the two-nucleon orbital angular momentum $L$,
spin $S$, and total angular momentum $J$ with projection $M$;
$\eta_S^{M_S}$ and $\zeta_T^{M_T}$ denote, respectively, the
eigenstates of the spin $S$ and isospin $T$ with projections $M_S$
and $M_T$. The deuteron $u(r)$ and $w(r)$, and $pp$ $\chi_L(r;k)$
radial wave functions are obtained from solutions of a Schr\"odinger
equation with nuclear and electromagnetic interactions, the latter
including corrections from vacuum polarization, magnetic moment,
two-photon exchange, and Darwin-Foldy terms. A discussion of the
interactions and radial wave functions is given in Secs.~III and IV
below. Here, it suffices to say that $\chi_L(r;k)$ behaves
asymptotically as
\begin{equation}
  \chi_L(r;k)_{\,\, {\scriptstyle \sim\atop\scriptstyle{r\to\infty}} }
      \cos\delta_L F_L(kr) +\sin\delta_L G_L(kr), \label{chiasy}
\end{equation}
where $\delta_L$ is the phase shift, and $F_L$ and $G_L$ are
the regular and irregular Coulomb functions.

The nuclear axial current operator consists of one- and two-body components
\begin{equation}
   {\bf A}_{a} = {\bf A}^{(1)}_a + {\bf A}^{(2)}_a,
\end{equation}
where $a=\pm$ is an isospin index, and
\begin{equation}
   {\bf A}^{(1)}_{\pm}=-g_A\sum_i{\bbox\sigma}_i\tau_{i,\pm}, \label{ax}
\end{equation}
\begin{equation}
   \tau_{i,\pm}=(\tau_{i,x}\pm{\rm i}\tau_{i,y})/2.
\end{equation}
The ratio of the axial to vector coupling constants, $g_A=G_A/G_V$, is
taken to be~\cite{Ade98} $1.2654\pm0.0042$ by averaging values obtained,
respectively, from the beta asymmetry in the decay of polarized neutrons
($1.2626\pm0.0033$)~\cite{PDG96,Abe97}, and $ft(n)$ and
$ft(0^+\rightarrow0^+)$, and $g_A=[2ft(0^+\rightarrow0^+)/ft(n)-1]/3
=1.2681\pm0.0033$~\cite{Ade98}. The form of the axial two-body current
operator depends on the dynamical model used to construct it. However,
the need for such a term is based on an analysis of tritium $\beta$ decay.
This evidence as well as the impact of the ambiguities associated with
the form of ${\bf A}^{(2)}$ on the $pp$ fusion cross section are
discussed below in Sec.~VI.

Selection rules for a vector operator restrict the sum over $L$ in the
initial capture state, Eq.~(\ref{scatt}), to the values $L=0$ and 2.
However, the $L=2$ contribution is negligible at very low energies.
Indeed, the initial $S$- and $D$-wave channel contributions to the
matrix element of the dominant ${\bf A}^{(1)}$ operator are proportional
to, respectively,
\begin{eqnarray}
  \int_0^\infty\!\! dr\,u(r)\chi_0(r;k) &\simeq&
         \int_0^\infty\!\! dr\,u(r)F_0(kr), \\
  \int_0^\infty\!\! dr\,w(r)\chi_2(r;k) &\simeq&
         \int_0^\infty\!\! dr\,w(r)F_2(kr), 
\end{eqnarray}
where the $\chi_L(r;k)$ radial wave functions have been replaced with
their asymptotic forms by setting $\delta_L \simeq 0$, which is
appropriate for the energy range under consideration here (a few keV).
It is then easily seen that
\begin{eqnarray}
  \frac{\int_0^\infty\!\! dr\,w(r)F_2(kr)}{\int_0^\infty\!\! dr\,u(r)F_0(kr)}
  \simeq && 2\sqrt{\left(1+\frac{1}{{\overline\eta}^2}\right)
                   \left(1+\frac{1}{4{\overline\eta}^2}\right)} \nonumber\\
  && \times\
  \frac{\int_0^\infty\!\! dr \sqrt{r}w(r)I_5(2\sqrt{\alpha m_p r})}
       {\int_0^\infty\!\! dr \sqrt{r}u(r)I_1(2\sqrt{\alpha m_p r})}, \nonumber\\
  &&
\end{eqnarray}
where ${\overline\eta}=\alpha/v_{\rm rel,n.r.}$, $I_L$ are modified
Bessel functions, and the asymptotic expressions, valid in the regime
where ${\overline\eta} >> kr$, have been used for the $F_L$~\cite{Abr70}.
The ratio above is found to be roughly 0.000\,13 in the limit
$v_{\rm rel,n.r.} \rightarrow 0$ (corresponding to ${\overline\eta}
\rightarrow\infty$).

Finally, the dependence of ${\bf A}_a$ upon the momentum transfer
${\bf q}=-{\bf p}_e-{\bf p}_\nu$, where ${\bf p}_e$ and ${\bf p}_\nu$
are the outgoing lepton momenta, is ignored in Eq.~(\ref{ax}), because
of the very low energies involved. At $E=0$ the kinetic energy available
to the final state is only about 420 keV, and the finite momentum
transfer correction to the matrix element of ${\bf A}^{(1)}$ for
$S$-wave capture can be estimated to be approximately $(q r_d)^2 \simeq
(0.0042)^2$, where $q\simeq 420$ keV and $r_d \simeq 2$ fm is the
root-mean-square radius of the deuteron---a tiny correction, indeed.

\section{$\protect\bbox{\lowercase{pp}}$ WAVE FUNCTION}
The low-energy $pp$ scattering is described by the radial Schr\"odinger
equation
\begin{equation}
   \left[\frac{d^2}{dr^2}+k^2-\frac{L(L+1)}{r^2}-m_pV(r)\right]
        \chi_L(r;k) = 0,            \label{SE}
\end{equation}
with $\chi_L(r;k)$ the radial wave function, $m_p$ the proton mass,
and $L$ the orbital angular momentum. The c.m.\ relative momentum $k$
is given by $k^2=m_pT_{\rm lab}/2$.
The boundary conditions for the wave function are
\[
   \chi_L(0;k)=0,
\]
\begin{equation}
   \chi_L(r;k)_{\,\, {\scriptstyle \sim\atop\scriptstyle{r\to\infty}} }
      F_L(kr)C_1 + G_L(kr)C_2,                     \label{CHIB}
\end{equation}
with $F_L$ and $G_L$ the standard regular and irregular Coulomb
functions~\cite{Abr70}.
The potential $V(r)$ can be divided into a long-range electromagnetic
part $V_{\rm EM}$ and a short-range nuclear part $V_N$.
The coefficients $C_1$ and $C_2$ contain all the necessary information
about the partial wave, which is usually expressed in terms of the
phase shift $\delta^C$:
\begin{equation}
   \tan\delta^C=C_2C_1^{-1}.               \label{DELTA}
\end{equation}
In a practical calculation, the Schr\"odinger equation (\ref{SE}) is
integrated out to some $r$, large in comparison with the range of the
short-range (i.e., nuclear) force. The numerical wave function is then
matched to the asymptotic form of Eq.~(\ref{CHIB}), and the
corresponding phase shift is defined to be the phase shift of the
nuclear force with respect to Coulomb wave functions. Following the
notation of Ref.~\cite{Ber88}, we have added the superscript $C$,
for Coulomb.

However, in reality the electromagnetic interaction in $pp$ scattering
is much more complicated than just the simple Coulomb interaction.
This leads to some practical problems in applying the scenario of
integrating the Schr\"odinger equation, matching to Coulomb functions,
and extracting the phase shift. This will be discussed below.

The full interaction, up to second order in the fine-structure
constant $\alpha\approx1/137$, is given by~\cite{Ber88,Wir95,Boh77}
\begin{equation}
   V_{\rm EM}(pp)=V_{C1}+V_{C2}+V_{\rm VP}+V_{\rm MM}+V_{\rm DF}, \label{VELM}
\end{equation}
where
\begin{equation}
  V_{C1}=\alpha' \frac{F_C(r)}{r},          \label{VC1}
\end{equation}
\begin{eqnarray}
  V_{C2} &=& -\frac{\alpha}{2m_p^2}\left[(\Delta+k^2)\frac{F_C(r)}{r}
              +\frac{F_C(r)}{r}(\Delta+k^2)\right]   \nonumber\\
  &\approx& -\frac{\alpha\alpha'}{m_p}\left[\frac{F_C(r)}{r}\right]^2,
                                                 \label{VC2}
\end{eqnarray}
\begin{equation}
  V_{\rm VP}=\frac{2\alpha}{3\pi}\,\frac{\alpha'}{r}I_{\rm VP}(r),
                                                 \label{VVP}
\end{equation}
\begin{eqnarray}
  V_{\rm MM} &=& -\frac{\alpha}{4m_p^2}\mu_p^2\left[
             \frac{2}{3}F_{\delta}(r)
             \bbox{\sigma}_1\!\cdot\!\bbox{\sigma}_2
            +\frac{F_t(r)}{r^3}S_{12}\right]     \nonumber\\
  && -\frac{\alpha}{2m_p^2}(4\mu_p-1)\frac{F_{ls}(r)}{r^3}
             {\bf L}\!\cdot\!{\bf S},            \label{VMM}
\end{eqnarray}
\begin{equation}
  V_{\rm DF}=-\frac{\alpha}{4m_p^2}F_{\delta}(r). \label{VDF}
\end{equation}
Here the $F_C(r)$, $F_{\delta}(r)$, $F_t(r)$, and $F_{ls}(r)$ are
functions representing the finite size of the nucleon charge
distribution.
In the limit of point nucleons, $F_C(r)=F_t(r)=F_{ls}(r)=1$, whereas
$F_{\delta}(r)=4\pi\delta^3({\bf r})$.
Their explicit $r$ dependence is given in Ref.~\cite{Wir95}.
The various contributions are described as follows.

The Coulomb potential $V_{C1}$ contains a well-known~\cite{Bre55}
energy dependence through $\alpha'=2k\alpha/(m_p v_{\rm lab})$.
However, at the extreme low energies of interest to astrophysical
calculations (a few keV), this energy dependence is negligible,
and we can set $\alpha'=\alpha$ for all practical purposes.

The two-photon-exchange interaction $V_{C2}$ behaves like $1/r^2$, and
so we immediately have the problem that, in principle, we have to
integrate out to infinitely large distances before we can match to
Coulomb functions.

The vacuum polarization potential $V_{\rm VP}$ describes the augmentation
of the photon propagator by an electron-positron pair. In the limit
of point protons [$F_C(r)=1$], the vacuum polarization integral is
given by~\cite{Ueh35}
\begin{equation}
   I_{\rm VP}(r)=\int_1^{\infty}\!\!dx\,e^{-2m_erx}
                 \left(1+\frac{1}{2x^2}\right)
                 \frac{(x^2-1)^{1/2}}{x^2},       \label{IVP}
\end{equation}
with $m_e=0.511$ MeV the electron mass. Including the finite-size
effect, we get the more complicated expression as given by Bohannon and
Heller~\cite{Boh77}, where the exponential is replaced by
\begin{eqnarray}
   e^{-2m_erx}\rightarrow && D^4(x)e^{-2m_erx} \nonumber\\
   &&-\left[D^4(x)+\frac{1}{2}\rho D^3(x)
                  +\frac{1}{8}(\rho+\rho^2)D^2(x) \right. \nonumber\\
   && \left.      +\frac{1}{48}(3\rho+3\rho^2+\rho^3)D(x)
                  \right]e^{-\rho},              \label{IVPFF}
\end{eqnarray}
where $\rho=br$, $D(x)=[1-(2m_ex/b)^2]^{-1}$, and $b=4.27$ fm$^{-1}$.
As a matter of fact, the simple multiplication of Eq.~(\ref{IVP}) with
$F_C(r)$ as an approximation to the inclusion of finite-size effects
(and which was adopted in Ref.~\cite{Wir95}) already closely resembles
the exact treatment (\ref{IVPFF}) where the finite-size effect is
properly folded into the integral.

The magnetic moment interaction $V_{\rm MM}$ arises as a consequence of
the nonvanishing value of the proton magnetic moment,
$\mu_p=2.792\,85\mu_0$, while the Darwin-Foldy term $V_{\rm DF}$ is a
short-range potential, describing the finite size of the proton, where
$F_{\delta}(r)\rightarrow4\pi\delta^3({\bf r})$ in the limit of
point protons.

Now that we have defined the full long-range electromagnetic interaction,
we can return to the question of how, in practice, to integrate the
Schr\"odinger equation and extract the phase shift. We will restrict
ourselves to $S$ waves, and so the tensor and spin-orbit terms in
the magnetic moment interaction vanish. It is convenient to define a
phase shift $\delta^V_W$, which is the phase shift of the solution of
the potential $W$ with respect to the solution with the potential $V$
as the interaction. The well-known application of this procedure is the
situation where $V$ is the Coulomb potential and $W$ is the Coulomb
plus nuclear potential, and the phase shift $\delta_N=\delta^C_{C+N}$
is obtained by matching the numerical solution to Coulomb wave functions
as in Eq.~(\ref{CHIB}).

If we are to include $V_{C2}$ in $W$, we run into the problem that we
have to integrate out to infinitely large distances before we can match
to Coulomb functions. However, by including the point-nucleon limit of
$V_{C2}$ also in $V$, the asymptotic wave function can be expressed in
terms of noninteger $L'$ Coulomb functions, where $L'$ satisfies
\begin{equation}
            L'(L'+1)=L(L+1)-\alpha\alpha'.
\end{equation}
The asymptotic behavior of the wave function is now given by
\begin{equation}
   \chi_L(r) \sim \widetilde{F}_L(kr)C_1 + \widetilde{G}_L(kr)C_2,
\end{equation}
with $\widetilde{F}_L(kr)=F_{L'}(kr)$ and similarly for
$\widetilde{G}_L(kr)$.
The advantage is clear immediately: we only have to integrate out to
distances large with respect to the nuclear interaction, which is only
about 20 fm. But we have to bear in mind that now the phase shift is
$\delta^{C1+C2}_{C1+C2+{\rm FS}+N}$, where it should be understood
implicitly that the superscript refers to the interactions in the
point-nucleon limit, while the interaction denoted by the subscript
includes also the (short-range) finite-size effects. To make this clear
explicitly, we have here separated off the finite-size effects by writing
them symbolically as being due to some short-range potential $V_{\rm FS}$.
In this notation, the phase shift with respect to Coulomb functions,
as defined in Eq.~(\ref{DELTA}), now reads
\begin{eqnarray}
   \delta^{C1} &=& \delta^{C1}_{C1+C2+{\rm FS}+N}
     = \delta^{C1+C2}_{C1+C2+{\rm FS}+N} + \delta^{C1}_{C1+C2} \nonumber\\
     &=& \delta^{C1+C2}_{C1+C2+{\rm FS}+N} + \rho_L,
\end{eqnarray}
where $\rho_L$ can be easily expressed in terms of the standard
Coulomb phase shift $\sigma_L$ as
\begin{equation}
    \rho_L=\sigma_{L'}-\sigma_L-(L'-L)\pi/2.
\end{equation}

The next step is to also include the vacuum polarization. The case
where we only have the Coulomb and vacuum polarization has been
discussed in detail by Durand~\cite{Dur57} and Heller~\cite{Hel60},
who derive expressions for the relevant asymptotic wave functions and
vacuum polarization phase shift $\tau_L\equiv\delta^{C1}_{C1+{\rm VP}}$.
Although the vacuum polarization potential exhibits an exponential
falloff, the small value of the electron mass means that the
Schr\"odinger equation has to be integrated out to several hundred
Fermi before the potential has dropped to sufficiently small values,
and it is only then that the numerical solution can be properly
matched to the asymptotic solution.

The presence of $V_{C2}$ considerably worsens the situation. Since
the $1/r^2$ behavior of $V_{C2}$ is of longer range than the exponential
decay of $V_{\rm VP}$, the Schr\"odinger equation has to be integrated out
to distances where $V_{\rm VP}$ is negligibly small as compared to $V_{C2}$.
It is only then that we can match the numerical solution to the proper
asymptotic solution and define the phase shift. Unfortunately, because
of the slow falloff of the vacuum polarization and the small magnitude
of the two-photon-exchange contribution, we now have to integrate out
to much larger distances. Even at a distance of 2000 fm the vacuum
polarization has only dropped to about 1\% of the two-photon exchange.

The scenario of getting the $pp$ wave function for a particular nuclear
interaction $V_N$ in the $S$ wave in the presence of the full
electromagnetic interaction $V_{\rm EM}$ is now as follows. We integrate
the Schr\"odinger equation out to a distance of 3000 fm, where the
numerical solution is matched to the electromagnetic wave functions
${\overline F}_0(kr)$ and ${\overline G}_0(kr)$. The latter are defined
to be the solutions of the Schr\"odinger equation in the presence
of the point-nucleon $C1+C2+{\rm VP}$ interaction. This procedure,
therefore, determines the phase shift $\delta^{\rm EM}$ of the nuclear
plus full electromagnetic interaction with respect to the point-nucleon
$C1+C2+{\rm VP}$ interaction.
It should be stressed that this phase shift is {\it not\/} the same
as the phase shift of the nuclear interaction in the presence of
only the Coulomb interaction ($\delta^C_{C+N}$).
The relation between these electromagnetic wave functions and the
standard Coulomb wave functions $F_0(kr)$ and $G_0(kr)$ is given by
\begin{equation}
    \left(\begin{array}{c}
           {\overline F}_0 \\ {\overline G}_0
          \end{array}\right)
    = \left(\begin{array}{rr}
            \cos(\rho_0+\tau'_0) & \sin(\rho_0+\tau'_0) \\
           -\sin(\rho_0+\tau'_0) & \cos(\rho_0+\tau'_0)
          \end{array}\right)
    \left(\begin{array}{c} F_0 \\ G_0 \end{array}\right), \label{PHIELM}
\end{equation}
with $\rho_0$ and $\tau'_0$ the two-photon-exchange and vacuum
polarization $S$-wave phase shifts, respectively. The prime in the
vacuum polarization is to indicate that this is the vacuum
polarization phase shift {\it in the presence of $V_{C2}$}, which
is slightly different from what is defined in Refs.~\cite{Dur57,Hel60}.
Note that the numerical wave function is now properly normalized as
in Eq.~(\ref{CHIB}), since $\delta^C=\delta^{\rm EM}+\rho_0+\tau'_0$.

It should be pointed out that at extreme low energies (a few keV),
$\delta^{\rm EM}$ is almost zero, $\rho_0$ is of the order of a few
times $10^{-4}$ deg, whereas $\tau'_0$ rapidly drops from about
$-10^{-2}$ deg at 10 keV to $-10^{-5}$ deg at 2 keV, and so
$\delta^C$ exhibits a change of sign and goes through zero as a function
of energy. Hence, it is not recommended to use the normalization as
advocated by Kamionkowski and Bahcall in Ref.~\cite{Kam94}, i.e.,
\begin{equation}
  \overline{\chi}_0(r;k)_{\,\, {\scriptstyle
                          \sim\atop\scriptstyle{r\to\infty}} }
  C_0 \left[ G_0(kr) +\cot\delta_0 \, F_0(kr) \right],   \label{BAHNORM}
\end{equation}
with $C_0$ the Gamow penetration factor. In their case~\cite{Kam94},
there is no problem (although $\delta^C$ is very small and
$\cot\delta^C$ becomes very large), because they did not include
$V_{C2}$.
Furthermore, with this normalization (\ref{BAHNORM}), the overlap
integral $\Lambda$, defined below, requires knowledge of the $pp$
scattering length $a_{pp}$, where the presence of $V_{C2}$ and
$V_{\rm VP}$ in the full electromagnetic interaction defines a rather
complicated effective-range function~\cite{Ber88}.
On the other hand, the normalization (\ref{CHIB}) advocated here
allows for an immediate substitution of the numerical wave function
(as obtained from solving the Schr\"odinger equation) into the
expression for $\Lambda$ as defined by Salpeter~\cite{Sal52},
without having to worry about a phase shift which goes through zero
at these extreme low energies and without having to define a
complicated effective-range function.

\section{DEUTERON WAVE FUNCTION}
The deuteron is the bound state of protons and neutrons in the coupled
$^3S_1$+$^3D_1$ two-nucleon system. For a given local $N\!N$ potential
$V({\bf r})$, the radial wave functions $u(r)$ and $w(r)$ for the
deuteron $S$ and $D$ states, respectively, can be obtained from the
coupled Schr\"odinger equation
\[
  \left[\frac{d^2}{dr^2} - \gamma^2 \right] u(r) =
  \overline{m} \left[ V_{00}(r)u(r) + V_{02}(r)w(r) \right],
\]
\begin{equation}
  \left[\frac{d^2}{dr^2} - \gamma^2 -\frac{6}{r^2} \right] w(r) =
  \overline{m} \left[ V_{20}(r)u(r) + V_{22}(r)w(r) \right], \label{DE}
\end{equation}
where $\overline{m}$ is twice the reduced mass of proton and neutron,
i.e.,
\begin{equation}
   \overline{m} \equiv \frac{2 m_p m_n}{m_p + m_n}.  \label{Mbar}
\end{equation}
All $N\!N$ potentials applied in this study use consistently the latest,
very accurate, values for nucleon masses~\cite{PDG96}, namely,
\begin{eqnarray}
        m_p & = & 938.272\,31 \mbox{ MeV,}\\
        m_n & = & 939.565\,63 \mbox{ MeV,}
\end{eqnarray}
implying
\begin{eqnarray}
        \overline{m} & = & 938.918\,52 \mbox{ MeV.}
\end{eqnarray}
For the $N\!N$ potential acting in particular partial waves,
we have introduced the convenient shorthand notation
$V_{00}(r)\equiv \langle^3S_1|V|^3S_1\rangle$,
$V_{02}(r)\equiv \langle^3S_1|V|^3D_1\rangle$, etc., where
$\langle{\bf\hat{r}}|^3S_1\rangle\equiv{\cal Y}^{1M}_{01}({\bf\hat{r}})$
and
$\langle{\bf\hat{r}}|^3D_1\rangle\equiv{\cal Y}^{1M}_{21}({\bf\hat{r}})$.
The quantity $\gamma=ik$ is discussed below.

The radial wave functions are properly normalized to unity,
\begin{equation}
   \int_0^\infty\!\! dr \left[ u^2(r) + w^2(r) \right] = 1.
\end{equation}
The asymptotic behavior of the wave functions for large values of $r$ is
\[
  u(r) \sim A_S e^{-\gamma r},
\]
\begin{equation}
  w(r) \sim A_D e^{-\gamma r}
      \left[1+\frac{3}{(\gamma r)}+\frac{3}{(\gamma r)^2}\right],
                                    \label{DAW}
\end{equation}
where $A_S$ and $A_D$ are known as the asymptotic $S$- and $D$-state
normalizations, respectively. In addition, one defines the
``$D/S$-state ratio'' $\eta\equiv A_D/A_S$.

Other deuteron parameters of interest are the quadrupole moment
\begin{equation}
  Q_d = \frac{1}{20} \int_0^\infty\!\! dr\,r^2 w(r)
        \left[ \sqrt{8} u(r) - w(r) \right],
\end{equation}
the root-mean-square or matter radius
\begin{equation}
  r_d = \frac12 \left\{\int_0^\infty\!\! dr\,r^2 \left[u^2(r) + w^2(r)
                \right]\right\}^{1/2},
\end{equation}
and the $D$-state probability
\begin{equation}
   P_D = \int_0^\infty\!\! dr\,w^2(r).
\end{equation}

Similar to scattering, the deuteron equation, Eq.~(\ref{DE}), is
solved numerically by integrating out to some large $r$ (25 fm in our
case) and matching the numerical waves to their asymptotic forms,
Eq.~(\ref{DAW}), producing $A_S$, $A_D$, and $\gamma$ from which the
predicted deuteron binding energy is extracted.

As mentioned, in the Schr\"odinger equation, Eq.~(\ref{DE}), the
interaction between the two nucleons is represented by a local
potential $V({\bf r})$, with ${\bf r} = {\bf r}_2 - {\bf r}_1$
the relative displacement between nucleons 1 and 2.
However, in general, the $N\!N$ potential $V$ is nonlocal, i.e.,
$V\equiv V({\bf r},{\bf r'})$, where ${\bf r}$ is the distance
between the two ingoing nucleons and ${\bf r'}$ the one between
the outgoing nucleons. A local potential can then be written as
$V({\bf r},{\bf r'})|_{\rm local}=\delta({\bf r}-{\bf r'}) V({\bf r})$.
For the more general case of a nonlocal potential, the coupled 
Schr\"odinger equation reads
\begin{eqnarray*}
  \left[\frac{d^2}{dr^2} - \gamma^2 \right] u(r) &=&
  \overline{m} \int_0^\infty\!\! dr'\,r r' \left[ V_{00}(r,r')u(r') \right. \\
  && \left. +V_{02}(r,r')w(r')\right],
\end{eqnarray*}
\begin{eqnarray}
  \left[\frac{d^2}{dr^2} - \gamma^2 - \frac{6}{r^2} \right] w(r) &=&
  \overline{m} \int_0^\infty\!\! dr'\,r r'
        \left[ V_{20}(r,r') u(r') \right. \nonumber\\
  && \left. +V_{22}(r,r')w(r')\right].                 \label{DENL}
\end{eqnarray}
This system of coupled integro-differential equations is then
solved by a combination of finite-difference, integral-discretization,
and matrix-inversion techniques.

Alternatively, one may consider the two-nucleon system in momentum
space, where the deuteron wave function is given by
\begin{equation}
  \Psi_d^M({\bf q}) = \left[
  \psi_0(q) {\cal Y}^{1M}_{01}({\bf \hat{q}}) +
  \psi_2(q) {\cal Y}^{1M}_{21}({\bf \hat{q}}) \right]
  \zeta_0^0,                                           \label{DWQ}
\end{equation}
with the normalization
\begin{equation}
  \int_0^\infty\!\! dq\,q^2 \left[ \psi_0^2(q) + \psi_2^2(q) \right]=1.
\end{equation}

The momentum-space Schr\"odinger equation that corresponds to
Eq.~(\ref{DENL}) consists of two coupled integral equations,
\begin{eqnarray*}
  \psi_0(q) &=& -\frac{\overline{m}}{\gamma^2+q^2}
                 \int_0^\infty\!\! dq'\,q'^2
                 \left[V_{00}(q,q')\psi_0(q') \right. \\
  && \left. + V_{02}(q,q')\psi_2(q') \right],
\end{eqnarray*}
\begin{eqnarray}
  \psi_2(q) &=& -\frac{\overline{m}}{\gamma^2+q^2}
                 \int_0^\infty\!\! dq'\,q'^2
                 \left[V_{20}(q,q')\psi_0(q') \right. \nonumber\\
  && \left. + V_{22}(q,q')\psi_2(q') \right].       \label{DEQ}
\end{eqnarray}
Considering a finite set of discrete arguments for the functions on the
left-hand side (LHS) and using the same set of momenta to discretize
the integrals on the RHS produces a matrix equation that is solved
easily by the matrix-inversion method~\cite{Haf70}.

The relevant Fourier transforms linking the configuration-space
and the momentum-space approaches are
\begin{eqnarray}
   V_{LL'}(q,q') &=& \frac{2}{\pi} \int_0^\infty\!\! dr\,r^2
                     \int_0^\infty\!\! dr'\,r^{'\,2}  \nonumber\\
   && \times\ j_L(qr) V_{LL'}(r,r') j_{L'}(q'r'),      \label{DFTV}
\end{eqnarray}
with $V_{LL'}(r,r')|_{\rm local} = \delta(r-r')V_{LL'}(r)/rr'$ if the
potential is local, and
\begin{equation}
  \frac{u_L(r)}{r} = \sqrt{\frac{2}{\pi}}
  \int_0^\infty\!\! dq\,q^2 j_L(qr) \psi_L(q),         \label{DFTW}
\end{equation}
with $u_0(r)\equiv u(r)$, $u_2(r)\equiv w(r)$, and $j_L$ the spherical
Bessel functions.

Since high reliability and precision is an important aspect of our present
investigation, we have calculated the deuteron wave functions for some
local potentials both ways: first, by solving Eq.~(\ref{DE}) directly
and, second, by solving Eq.~(\ref{DEQ}) by matrix inversion and then
performing the transformation, Eq.~(\ref{DFTW}), numerically.
We find agreement between the resulting deuteron waves to at least
six significant digits for any $r$ in the range 0.05--14 fm.
This establishes the reliability of our numerical methods.
It also implies that in cases where we use the momentum-space approach
and Eq.~(\ref{DFTW}), as for the nonlocal potentials, our deuteron
waves are of the highest numerical precision.

The deuteron is a pole in the $S$ matrix at $k=i\gamma$.
The relativistic relationship between $\gamma$ and the deuteron
binding energy $B_d$ is given by~\cite{comm.}
\begin{equation}
   \sqrt{s}=m_d=m_p + m_n - B_d = \sqrt{m_p^2-\gamma^2} +
            \sqrt{m_n^2-\gamma^2},  \label{DGAMM}
\end{equation}
where $m_d$ denotes the deuteron rest mass.
Notice that this equation determines {\it the correct empirical
$\gamma$}, since nature is relativistic.
We note that in $N\!N$ scattering we use the relativistic relationship
between $k$ and $T_{\rm lab}$, which implies that the c.m.\ kinetic
energy $T$ is related to $k$ according to
\begin{equation}
    \sqrt{s}=m_p + m_n +T = \sqrt{m_p^2+k^2} + \sqrt{m_n^2+k^2}.
\end{equation}
Thus, consistency with the scattering problem requires the use of
Eq.~(\ref{DGAMM}) to determine $\gamma$.
The formal solution of Eq.~(\ref{DGAMM}) is
\begin{equation}
   \gamma^2 = \left[4m_p^2 m_n^2 - (m_d^2-m_p^2-m_n^2)^2\right]/4m_d^2,
              \label{GAMMASS}
\end{equation}
and, using $B_d=2.224\,575$ MeV and $\hbar c = 197.327\,053$ MeV fm,
the accurate numerical value for $\gamma$ comes out to be
\begin{equation}
   \gamma = 0.231\,538\,0 \mbox{ fm}^{-1}.  \label{DGAMMA}
\end{equation}
To obtain some pedagogical insight into $\gamma^2$,
one may rewrite Eq.~(\ref{GAMMASS}) in factorized form
\begin{eqnarray}
   4m_d^2\gamma^2 &=& \left[(m_n+m_p)^2-m_d^2\right]
                      \left[m_d^2-(m_n-m_p)^2\right]  \nonumber\\
   &=& B_d(4m-B_d)(m_d^2-\delta m^2),
\end{eqnarray}
where we introduce the average nucleon mass,
\begin{equation}
   m \equiv \frac{m_p+m_n}{2}=938.918\,97 \mbox{ MeV},
\end{equation}
and the nucleon mass difference $\delta m \equiv m_n-m_p=1.293\,32$ MeV,
and used $m_d=2m-B_d$. {}From this we get
\begin{equation}
   \gamma^2=mB_d\left(1-\frac{B_d}{4m}\right)
                \left(1-\frac{\delta m^2}{m_d^2}\right),
\end{equation}
and rewriting twice the reduced nucleon mass [cf.\ Eq.~(\ref{Mbar})]
in terms of the average mass
\begin{equation}
   \overline{m}=m\left(1-\frac{\delta m^2}{4m^2}\right),
\end{equation}
we finally obtain
\begin{eqnarray}
    \gamma^2 &=& \overline{m}B_d\left(1-\frac{B_d}{4m}\right)
             \frac{1-\delta m^2/m_d^2}{1-\delta m^2/(4m^2)} \nonumber\\
    &\approx& \overline{m}B_d\left[1-\frac{B_d}{4m}
                \left(1-\frac{\delta m^2}{m^2}\right) \right]  \nonumber\\
    &\approx& \overline{m}B_d \left(1-\frac{B_d}{4m}\right).
              \label{gamma.app}
\end{eqnarray}
The approximations involved in Eq.~(\ref{gamma.app}) are good to 1 part
in $10^9$. Therefore, this equation reproduces the exact value for
$\gamma$ to all digits given in Eq.~(\ref{DGAMMA}).
One can now identify the term $\overline{m}B_d$ as the nonrelativistic
approximation to $\gamma^2$ and the factor $(1-B_d/4m)$ as the
essential relativistic correction. In most calculations of the past,
the nonrelativistic $\gamma$ was used,
$\gamma_{\rm nr}\equiv \sqrt{\overline{m} B_d}=0.231\,606\,6$ fm$^{-1}$.
The difference between $\gamma_{\rm nr}$ and the correct $\gamma$,
Eq.~(\ref{DGAMMA}), leads to a small difference (0.03\%) in the
overlap integral $\Lambda^2$ (see below). Although the difference is
rather small, we believe one should use the relativistically correct
value, Eq.~(\ref{DGAMMA}).

Besides the strong interaction, there is also a nonvanishing
electromagnetic interaction between protons and neutrons that can be
written as~\cite{Wir95}
\begin{equation}
   V_{\rm EM}(np) = V_{C1}(np) + V_{\rm MM}(np),            \label{VEMNP}
\end{equation}
where
\[
   V_{C1}(np) = \alpha\beta_n \frac{F_{np}(r)}{r},
\]
\begin{eqnarray}
   V_{\rm MM}(np)&=&-\frac{\alpha}{4m_pm_n}\mu_p\mu_n\left[\frac{2}{3}
           F_{\delta}(r) \bbox{\sigma}_1\!\cdot\!\bbox{\sigma}_2
           +\frac{F_t(r)}{r^3}S_{12}\right]       \nonumber\\
        &&  -\frac{\alpha}{m_p\overline{m}}\mu_n\frac{F_{ls}(r)}{r^3}
             ({\bf L}\!\cdot\!{\bf S}
             +{\bf L}\!\cdot\!{\bf A}).
\end{eqnarray}
Here ${\bf A}=(\bbox{\sigma}_1-\bbox{\sigma}_2)/2$, and
$F_{np}(r)$ is a short-range function representing the finite size
of the neutron charge distribution (for details, see Ref.~\cite{Wir95}).
Because the $S$-wave expectation values for the tensor and spin-orbit
operators vanish, the long-range $1/r^3$ parts do not contribute for
$L=0$. For $L\neq0$, we make the approximation that
$\delta^{\rm EM}_{\rm EM+N}\approx\delta_N$
(or $S^{\rm EM}_{\rm EM+N}\approx S_N$ in terms of the $S$ matrix).
This means that in our calculations the asymptotic behavior of the
deuteron wave functions still satisfies Eqs.~(\ref{DAW}) and (\ref{DEQ}).

The interaction (\ref{VEMNP}) is included only in the case of the
Argonne AV18 $N\!N$ potential~\cite{Wir95} where it contributes 18 keV
to the deuteron binding energy, mostly from the magnetic moment part
$V_{\rm MM}(np)$ of the interaction.
One would expect that the inner part of the deuteron wave function is
affected by the inclusion or omission of $V_{\rm EM}(np)$ (the outer
part is essentially insensitive since it is ruled by $\gamma$ which is
identical for all potentials). Fortunately, it turns out that the
quantitative effect is very small, as will be demonstrated below.
Thus, also models that do not include the electromagnetic interaction
between protons and neutrons can be considered as sufficiently reliable
for our study.

Since all $N\!N$ potentials are fitted to the value of $\gamma$ given
in Eq.~(\ref{DGAMMA}), they all accurately describe the empirical
deuteron binding energy, $B_d=2.224\,575(9)$ MeV~\cite{Leu82},
via the relativistic relation Eq.~(\ref{DGAMM}).
The other deuteron parameters, as well as the $^3S_1$ scattering length
$a_t$ and effective range $r_t$, are listed in Table~\ref{tabdeutpam}.
Predictions are given for the five high-precision $N\!N$ potentials
that we focus on, namely, AV18~\cite{Wir95}, CD-Bonn~\cite{Mac96},
Nijm-I~\cite{Sto94}, Nijm-II~\cite{Sto94}, and Reid93~\cite{Sto94}.
Notice that not all quantities in Table~\ref{tabdeutpam} are independent.
For example, the deuteron effective range $\rho_d\equiv\rho(-B_d,-B_d)$
is related to $A_S$, $\eta$, and $\gamma$ by
\begin{equation}
   A_S^2(1+\eta^2)=\frac{2\gamma}{1-\gamma\rho_d}.
\end{equation}
For our present investigation, essentially only $A_S$ is of relevance
(besides $\gamma$). However, $A_S$ (and $\rho_d$) cannot be measured
directly. The empirical information given in the last column of
Table~\ref{tabdeutpam} on $A_S$ and $\rho_d$ are model-dependent
extrapolations of low-energy data. Therefore, to trust the predictions
for $A_S$ by our $N\!N$ potentials, it is important that these models
reproduce accurately all measured low-energy data, which is confirmed
by Table~\ref{tabdeutpam}.
The only exceptions are the deuteron matter radius $r_d$ and the
quadrupole moment $Q_d$, which are both underpredicted by all potential
models. There are, however, meson-exchange current contributions and
relativistic corrections for $r_d$ and $Q_d$ which may make up for
the discrepancies~\cite{Fri97,Buc96}.
The $D$-state probability, $P_D$, that is listed in the bottom row of
Table~\ref{tabdeutpam}, is not an observable. It is, however, an
interesting theoretical quantity in studies of the nuclear force.
The lower value for $P_D$ predicted by CD-Bonn is a reflection of the
nonlocal nature of this potential which is based upon relativistic
meson field theory~\cite{Mac96,Mac89}.
Meson-exchange Feynman diagrams are, in general, nonlocal expressions
that are represented in momentum space in analytic form.

Finally, in Fig.~\ref{figdeut}, we display the deuteron wave functions
produced by the five $N\!N$ potential models. Major differences are,
again, related to whether the models are local or nonlocal.
While the central potentials of AV18, Nijm-II, and Reid93 are stricly
local, the Nijm-I central force includes momentum-dependent terms
which give rise to nonlocal structures in the equivalent
configuration-space potential. This affects the deuteron $S$ wave
and is the reason why the $u(r)$ generated by CD-Bonn and Nijm-I
are so similar (large solid and dashed curves in Fig.~\ref{figdeut})
and differ from the other three potentials.
The Nijm-I tensor potential is strictly local, similar to AV18,
Nijm-II, and Reid93, which explains why these four potentials generate
very similar $D$ waves. The CD-Bonn tensor potential is nonlocal.

\section{AXIAL ONE-BODY CURRENT CONTRIBUTION}
Using the wave functions as defined in Eqs.~(\ref{DW}) and (\ref{scatt})
and ignoring the $D$-wave contribution in the initial scattering state,
we find that the matrix element of the (dominant) one-body part of
the axial current is given by
\begin{eqnarray}
  \langle d,M | A^{(1)}_\mu | pp \rangle &=&
       \delta_{M,\mu}\sqrt{16\pi} g_A \frac{e^{{\rm i}\delta_0}}{k}
       \int_0^\infty\!\! dr u(r) \chi_0(r;k)           \nonumber\\
  &\equiv& \sqrt{\frac{32\pi}{\gamma^3}}g_AC_0\Lambda(E),
\end{eqnarray}
where $A^{(1)}_{\mu=\pm 1,0}$ are the spherical components of
${\bf A}^{(1)}$, $A_{\mu=\pm}=\mp(A_x \pm{\rm i}A_y)/\sqrt{2}$ and
$A_{\mu=0}=A_z$, $C_0$ is the Gamow penetration factor, and the overlap
integral is conventionally defined as~\cite{Sal52}
\begin{equation}
   \Lambda(E) = \left(\gamma^3/2\right)^{1/2}
                \frac{e^{{\rm i}\delta_0}}{C_0k}
                \int_0^{\infty}\!\!dr\,u(r) \chi_0(r;k).  \label{LAMPRACT}
\end{equation}
The constant $\gamma$ is defined in Eq.~(\ref{DGAMMA}), and the wave
function $\chi_0$ is normalized as in Eq.~(\ref{chiasy}). Because the
solar fusion reaction actually occurs at energies of only a few keV,
the phase shift $\delta_0$ is extremely small, and so the exponential
$e^{{\rm i}\delta_0}$ can conveniently be approximated by unity.
Note that when we adopt the normalization as advocated by Kamionkowski
and Bahcall~\cite{Kam94}, Eq.~(\ref{BAHNORM}), we find
\begin{equation}
    \Lambda(E) = \left(a_{pp}^2\gamma^3/2\right)^{1/2}
                 \int_0^{\infty}\!\!dr\,u(r) \overline{\chi}_0(r;k), \label{lkb}
\end{equation}
where the scattering length $a_{pp}$ is defined as
\begin{equation}
    -\,\frac{1}{a_{pp}}=\lim_{k\rightarrow0}C_0^2 k \cot\delta_0.
\end{equation}
Equation~(\ref{lkb}) coincides with the definition of the overlap
integral given by Kamionkowski and Bahcall~\cite{Kam94}. However, as
stated in our discussion on the $pp$ wave function, it is not at all
trivial to calculate the correct scattering length $a_{pp}$ when
electromagnetic interactions other than the point-particle Coulomb
interaction are present.

In the following, we will present our results for the overlap integral
using realistic $pp$ and deuteron wave functions. By realistic we
mean that these wave functions were obtained by solving the scattering
and bound-state equations using the recent high-precision $N\!N$
potential models, the parameters of which were fitted to give an almost
optimal description of the $N\!N$ scattering data up to laboratory energies
of 350 MeV (i.e., $\chi^2/{\rm data}\approx1$). The five $N\!N$ models
we consider consist of the AV18 Argonne model~\cite{Wir95}, the CD-Bonn
model~\cite{Mac96}, two Nijmegen models, Nijm-I and Nijm-II~\cite{Sto94},
and a regularized update of the Reid soft-core potential~\cite{Sto94}.
The AV18 potential was fitted including all finite-size effects in
the full electromagnetic potential of Eq.(\ref{VELM}), whereas the
other four potentials used the point-particle approximation, i.e.,
$F_C(r)=F_t(r)=F_{ls}(r)=1$ and $F_{\delta}(r>0)=0$.
Furthermore, the AV18 potential is the only model which includes the
electromagnetic interaction (\ref{VEMNP}) also in the deuteron.

In Table~\ref{tablamvnn} we show the results for $\Lambda^2(E_{\rm lab})$
($E=E_{\rm lab}/2$) as calculated from Eq.~(\ref{LAMPRACT}).
The integral was cut off at $r=50$ fm, which is valid since beyond this
distance the deuteron wave function has become extremely small, and so
the contribution to the overlap integral becomes negligible. The results
are shown for laboratory kinetic energies of 5, 4, 3, and 2 keV, which are
extrapolated to define the result at zero energy. For each model we
use the deuteron and $pp$ scattering wave functions of that particular
model. The dependence on the particular $N\!N$ model is found to be
rather small. Taking the average over all five models we find
$\Lambda^2(0)=6.975\pm0.010$.
Leaving out the CD-Bonn model, which is quite different from the
other models in that it is the only model with nonlocal tensor
interactions, we find an even smaller model dependence with
$\Lambda^2(0)=6.970\pm0.005$.

We again want to stress that these $N\!N$ models were fitted
including the full electromagnetic potential, and so the wave
functions have to be calculated in the presence of this same
electromagnetic interaction. Truncating it, for example by only
including the standard Coulomb interaction, will modify the
wave function and, hence, the overlap integral.
In Table~\ref{tabtrunc} we show the effect on $\Lambda^2(E)$ for
different truncations of the electromagnetic part of the interaction.
For the nuclear interaction we take the AV18 potential as an example.
The other models show a similar trend. We consider four different
truncations of the electromagnetic interaction, all for point-particle
protons. The effect of $V_{C2}$ is seen to be rather small:
neglecting it increases $\Lambda^2(0)$ by only 0.0035, which is a
0.05\% effect. The proper inclusion of the vacuum polarization is
much more important: neglecting it causes an almost 1\% increase.

Finally, for the AV18 potential we can also study the finite-size
effects and the effect of $V_{\rm EM}(np)$ in the deuteron calculation.
Neglecting the finite-size effects underestimates $\Lambda^2(0)$ by
only 0.08\%, as shown in the table. Simply removing $V_{\rm EM}(np)$
changes the binding energy to $B_d({\rm trunc})=-2.242\,227$ MeV, and
hence the asymptotic behavior of the deuteron wave function. The
consequence of this is that $\Lambda^2(0)$ increases by 0.03, almost
a 0.5\% effect. However, if we first refit the binding energy
[i.e., make a modified AV18 potential which does not include
$V_{\rm EM}(np)$, but which does have the proper asymptotic deuteron
wave function], then the difference in $\Lambda^2(0)$ is only 0.001.
Hence, the inclusion of $V_{\rm EM}(np)$ under the restriction that the
potential model correctly fits the experimental binding energy has
only a small effect on the overlap integral, as we alluded to earlier.

\section{BEYOND THE AXIAL ONE-BODY CURRENT CONTRIBUTION}
In this section we review the procedure leading to the experimental
determination of the Gamow-Teller (GT) matrix element in tritium
$\beta$ decay, and demonstrate the inability of calculations based on
axial one-body currents and realistic wave functions from modern
interactions to correctly predict this value. After a brief discussion
of the axial two-body current operators, we address the issue of their
model dependence by adopting the phenomenological approach of
constraining them to reproduce the experimental value of the $^3$H GT
matrix element. We then calculate these two-body current contributions
to the $pp$ weak capture, examining in particular the question of how
their associated uncertainties affect the $pp$ cross section.

\subsection{Tritium $\protect\bbox{\beta}$ decay}
Evidence for the presence of axial two-body current contributions
to weak transitions comes from the $\beta$ decay of tritium.
Its half-life can be expressed as
\begin{equation}
  (1+\delta_R)t = \frac{ K/G_V^2}{f_V \langle{\bf F} \rangle^2 +
                 f_A \, g_A^2 \langle{\bf GT} \rangle^2},
\end{equation}
where $\delta_R$=1.9\% is the so-called outer radiative correction,
$t$ is the half-life, and $f_V$ and $f_A$ are Fermi functions calculated
by Towner, as reported by Simpson~\cite{Sim87}, to have the values
$2.8355\times 10^{-6}$ and $2.8505\times 10^{-6}$, respectively.
The experimental value for the combination $K/G_V^2$ is
($6146.6 \pm 0.6$)~s, as obtained by Hardy {\it et al.}~\cite{Har90}.
This value is actually 0.15\% larger than that used by
Simpson~\cite{Sim87}, ($6137.2 \pm 3.6$)~s, in his $^3$H $\beta$-decay
analysis. Finally, $\langle{\bf F}\rangle$ and $\langle{\bf GT}\rangle$
denote the reduced matrix element of the Fermi and GT operators, which
in the one-body limit are given by, respectively
\begin{equation}
  \langle{\bf F}\rangle = \langle {^3{\rm He}}\mid\mid \sum_i \tau_{i,+} 
         \mid\mid {^3{\rm H}}\rangle,
\end{equation}
\begin{equation}
  \langle{\bf GT}\rangle = \langle {^3{\rm He} }\mid\mid \sum_i 
         {\bbox\sigma}_i \tau_{i,+} \mid\mid {^3{\rm H}}\rangle.
\end{equation}

Simpson~\cite{Sim87} reports the experimental value ($1134.6 \pm 3.1$)~s
for the combination $(1+\delta_R) t f_V$. In order to extract a value
for the tritium GT matrix element, it is necessary to calculate the
Fermi matrix element. If the trinucleons were pure total $T=1/2$,
$M_T=\pm 1/2$ states, then the Fermi matrix element would just be one.
However, charge-symmetry breaking (CSB) and charge-independence breaking
(CIB) and, more importantly, electromagnetic effects in the nuclear
interaction lead to a small correction. In the present study, such a
correction is calculated using $^3$H and $^3$He wave functions obtained
with the correlated-hyperspherical-harmonic (CHH) method~\cite{Kie95}
from the AV18 two-nucleon interaction (including electromagnetic terms)
and the Urbana UIX three-nucleon interaction~\cite{Pud95}. We find,
neglecting isospin admixtures $T \ge 3/2$ (the probability of $T=3/2$
components in $^3$He has been estimated to be about 0.0016\%),
\begin{equation}
  \langle{\bf F}\rangle^2 \equiv 1-\epsilon =0.9987.
\end{equation}
The present value for $\epsilon$ is about twice that obtained by 
Saito {\it et al.}~\cite{Sai90} in a (converged) Faddeev calculation
based on the older Argonne $v_{14}$ two-nucleon~\cite{Wir84} and
Tucson-Melbourne (TM) three-nucleon~\cite{Coo79} interactions and
phenomenological CSB and CIB terms constrained to reproduce the
observed mass difference in $^3$H and $^3$He. However, the individual
binding energies are underpredicted by this Hamiltonian model by
about 3\%. In contrast, the present AV18/UIX CHH wave functions
reproduce the experimental binding energies of both systems within
less than 10 keV (incidentally, the variational CHH and ``exact''
Faddeev~\cite{Nog97} and Green's function Monte Carlo~\cite{Pud97}
methods produce trinucleon binding energies all within a few keV of
each other). It is unclear at this point whether the difference in
$\epsilon$ values calculated here and in Ref.~\cite{Sai90} is to be
ascribed to binding energy effects or to differences in the treatment
of the electromagnetic, CSB, and CIB interactions (or both).
We note that Simpson uses the value $\epsilon=0.0006$ in line with
the estimate of Saito {\it et al.}

Using the measured half-life, and the values $K/G_V^2=(6146.6\pm0.6)$~s,
$f_A/f_V=1.005\,29$, $\langle{\bf F}\rangle^2=0.9987$, and
$g_A=1.2654 \pm 0.0042$, the ``experimental'' GT matrix element is obtained:
\begin{equation}
   \langle{\bf GT}\rangle \mid_{\rm exp} =\sqrt{3}(0.957\pm0.003),
\end{equation}
where the $\sqrt{3}$ is from a Clebsch-Gordan coefficient.

The experimental GT matrix element is compared with predictions
from a number of modern Hamiltonians with various combinations
of realistic two- and three-nucleon interactions in Table~\ref{tabgt}. 
We also give in Table~\ref{tabpro} the calculated percent probabilities
of the $S$-, $S^\prime$-, $P$- and $D$-wave components in the $^3$H wave
function~\cite{Kie95,Nog97}.
A few comments are in order. First, the model Hamiltonians with
the TM three-nucleon interaction are all designed to reproduce the
experimental $^3$H binding energy in Faddeev calculations by adjusting
the cutoff mass in the TM force~\cite{Nog97}. As already pointed out,
the two-nucleon interactions employed in the present work are
of high precision, and produce fits to $pp$ and $np$ scattering data
up to laboratory energies of 350 MeV with a $\chi^2$ per datum in the range
1.03--1.09.  

Second, in Table~\ref{tabgt} we also quote the results obtained
using the relation
\begin{equation}
  \langle{\bf GT}\rangle \simeq \sqrt{3} (P_S+P_D/3-P_{S^\prime}/3),
                                    \label{ppp}
\end{equation}
where $P_S$, $P_D$, and $P_{S^\prime}$ are the probabilities of the
$S$-, $D$- and $S^\prime$-wave components in the $^3$H state.
Use of such a relation implicitly assumes isospin symmetry---namely,
that $^3$H and $^3$He form an isodoublet---and also ignores the
contribution of $P$-wave components.
However, corrections to Eq.~(\ref{ppp}) appear to be very small, a few
parts in a thousand.  

Third, the results listed in Table~\ref{tabgt} indicate that modern
interactions lead to predictions for the GT matrix element of tritium
in the range $\sqrt{3}\times (0.923-0.937)$, and therefore to an
underestimate of the experimental value ranging, in relative terms,
from 2.1\% for CD-Bonn/TM to 3.7\% for AV18/UIX.

\subsection{Axial two-body current model}
For the axial two-body current operator we use a slightly expanded version
of the conventional $\pi$- and $\rho$-meson exchange model first described
by Chemtob and Rho~\cite{Che71}. These are two-body currents associated
with excitation of intermediate $\Delta$ resonances by $\pi$ and
$\rho$ exchanges, the $\pi \rho$ mechanism, and the contact $\pi N\!N$
and $\rho N\!N$ interactions. In the tables, these operators are denoted,
respectively, as $\Delta \pi$, $\Delta \rho$, $\pi \rho$, $\pi S$, and
$\rho S$. Explicit expressions for them are listed in the Appendix for
completeness.
Here we only note that (i) the (nonlocal) momentum-dependent terms in
the $\pi$, $\rho$, and $\pi \rho$ operators are retained in contrast
to Ref.~\cite{Sai90}; (ii) monopole form factors are included at the
$\pi N\!N$ and $\rho N\!N$ vertices with cutoff masses $\Lambda_\pi$ and
$\Lambda_\rho$, respectively; and (iii) there is significant uncertainty in
the leading $\Delta \pi$ and $\Delta \rho$ contributions, since the
$N$ to $\Delta$ transition axial coupling is not known~\cite{Sch92}.
In the model adopted here, the latter is related within the quark model
to the nucleon $g_A$, namely, $g_{N\Delta}=(6\sqrt{2}/5)g_A$.

The present approach consists in using the simplest possible two-body
operators that give an adequate description of the longest-range
mechanisms and of adjusting the cutoff masses {\it within a given
Hamiltonian model\/} so as to reproduce the experimental $^3$H GT
matrix element. The contributions due to exchanges of heavier mesons,
such as the $A_1$~\cite{Cie84,Tow87}, or renormalizations effects,
arising from $\Delta$-isobar admixtures in the nuclear wave
functions~\cite{Sch92}, are neglected. However, in the next subsection
it is argued that these approximations are not expected to impact in
any significant way on the theoretical predictions for the $pp$ weak
capture cross section once the two-body current model is constrained
to fit the GT matrix element of tritium.

\subsection{Axial two-body current contributions to the
            $\protect\bbox{pp}$ capture and $^{\protect\bf 3}$H
            GT matrix element}
In Table~\ref{tabgt2} we quote the contributions to the GT matrix
element obtained with the CHH AV18/UIX trinucleon wave functions from
the individual components of the axial current operators listed in
the Appendix. The small differences between the present results and
those reported in Ref.~\cite{Car91} are due to the slightly different
values used for $\Lambda_\pi$ ($\Lambda_\pi=4.80$ fm$^{-1}$ in the
present work versus $\Lambda_\pi=4.65$ fm$^{-1}$ in Ref.~\cite{Car91})
and, presumably to a lesser extent, to the fact the older calculations
were based on a different Hamiltonian model, consisting of the Argonne
$v_{14}$ two-nucleon and Urbana-VIII three-nucleon interactions,
which, however, did reproduce the experimental binding energies of the
trinucleons in a 34-channel Faddeev calculation~\cite{Che86}.
The cumulative value for the calculated GT matrix element is
$\sqrt{3} \times 0.964$, about 0.7\% larger than experiment.
A slight adjustment in the cutoff masses $\Lambda_\pi$ and $\Lambda_\rho$
or $N\Delta$ axial coupling (or both) is thus required to bring theory
and experiment into perfect agreement.
We will return to this point later, in Sec.~VII.

To test the model dependence, we have calculated the leading $\Delta\pi$
contribution with 42-channel Faddeev wave functions obtained from the
Hamiltonian models discussed earlier, and the results are listed in
Table~\ref{tabgt}. Both the one-body and $\Delta\pi$ contributions show
a strong correlation with the $D$-state probability in the trinucleon
wave functions, which is obviously related to the deuteron $D$-state
probability predicted by the underlying two-nucleon interaction, as is
evident from Tables~\ref{tabdeutpam} and~\ref{tabpro}. This correlation
is a direct consequence of the dominant contributions due to $T=1$ $^1S_0
\rightleftharpoons T=0$ $^3S_1$-$^3D_1$ ($T=0$ $^3D_1$) transitions
for the one-body ($\Delta\pi$) component.
This has been verified explicitly by including only the above channels
in the Faddeev evaluation of the GT matrix element. As a result, the
sum of the one-body and $\Delta \pi$ contributions turns out to be
essentially model independent, as indicated in Table~\ref{tabgt}.
Such a conclusion is also expected to hold when the remaining two-body
contributions are included. Thus, to reproduce the experimental GT
matrix element, a single adjustment of the cutoff masses $\Lambda_\pi$
and $\Lambda_\rho$ or $g_{N\Delta}$ in the axial two-body current
operators should suffice for all Hamiltonian models considered.

We now turn to the $pp$ capture. We only quote results, presented in
Table~\ref{tablpp}, corresponding to the AV18 and CD-Bonn interactions.
The values calculated with these two models, which give the two extremes
for the one-body contribution, 6.966 and 6.992, respectively, at zero
energy, are within less than 0.2\% when all two-body current
contributions are included. Thus, the two-body part of the axial current
leads to an increase of the AV18 and CD-Bonn one-body results, amounting,
respectively, to 1.6\% and 1.1\%, consistently with the findings of the
earlier study~\cite{Car91}. 

Having demonstrated the model independence of theoretical predictions
for the GT matrix element and $pp$ weak capture cross section, we now
want to address the issue of how ambiguities in the axial two-body
currents might affect this conclusion. To this end, it is useful to
decompose the GT matrix element as
\begin{eqnarray}
   \langle{^3{\rm He}}\mid && \sum_{i<j}O_{z,+}(ij)\mid {^3{\rm H}}\rangle
                              \nonumber\\
   = && \langle{^3{\rm He}}\mid\sum_{i<j}O_{z,+}(ij)P^\tau_1(ij)
                            \mid{^3{\rm H}}\rangle      \nonumber\\
   && +\langle{^3{\rm He}}\mid\sum_{i<j}O_{z,+}(ij)P^\tau_0(ij)
                            \mid{^3{\rm H}}\rangle,         \label{gt01}
\end{eqnarray}
where $O_{z,+}$ is the $z$ component of any axial two-body current
operator, and $P^\tau_{0,1}$ are projection operators over $T=0$ and 1
two-nucleon states:
\begin{equation}
   P^\tau_0(ij)+P^\tau_1(ij)=1,
\end{equation}
\begin{equation}
   P^\tau_1(ij)=\frac{3+{\bbox\tau}_i \cdot {\bbox\tau}_j}{4}.
\end{equation}
In Eq.~(\ref{gt01}) most of the $T=0$ ($T=1$) contribution is coming
from conversion of a $pn$ $T,S=0,1$ ($nn$ $T,S=1,0$) pair in $^3$H to a
$pp$ $T,S=1,0$ ($pn$ $T,S=0,1$) pair in $^3$He, for example,
\begin{eqnarray}
   \langle{^3{\rm He}} && \mid\sum_{i<j}O_{z,+}(ij)P^\tau_1(ij)
           \mid{^3{\rm H}}\rangle       \nonumber\\
   \simeq && \langle{^3{\rm He}}\mid\sum_{i<j}P^\tau_0(ij)
           O_{z,+}(ij)P^\tau_1(ij)\mid{^3{\rm H}}\rangle,
\end{eqnarray}
since the numbers of $T,S=0,0$ and $T,S=1,1$ pairs in the trinucleons
are much smaller than those with $T,S=0,1$ and $T,S=1,0$~\cite{For96}.
It is now easy to see that, if (neglecting isospin-symmetry breaking
corrections) $\mid {^3{\rm He}}\rangle = Q \mid {^3{\rm H}}\rangle$,
where $Q \equiv \tau_{1,x} \tau_{2,x} \tau_{3,x}$ is the isospin-flip
operator, then
\begin{eqnarray}
   \langle{^3{\rm He}} && \mid\sum_{i<j}P^\tau_0(ij)O_{z,+}(ij)P^\tau_1(ij)
           \mid{^3{\rm H}}\rangle        \nonumber\\
   = && \langle{^3{\rm He}}\mid\sum_{i<j}P^\tau_1(ij)
           O_{z,+}(ij)P^\tau_0(ij)\mid{^3{\rm H}}\rangle,
\end{eqnarray}
since the matrix element is real, $Q$ commutes with $P^\tau_T$, $Q^2=1$,
and $O_\pm^\dagger=O_\mp$. Thus, the $T=0$ and $T=1$ contributions
in Eq.~(\ref{gt01}) are expected to be of about the same size. This
can be seen from Table~\ref{tabgt2}, where the sum of the $T=0$ and 1
and $T=1$ alone contributions to the GT matrix element from the
individual components of the two-body operators are listed.

It is interesting to define the two-body densities:
\begin{equation}
   \rho_O(x;{\rm GT}) = \langle {^3{\rm He}} \mid 
          \sum_{i<j}\delta(x-r_{ij})O_{z,+}(ij)P^\tau_0(ij)
          \mid{^3{\rm H}}\rangle,                \label{p00}
\end{equation}
\begin{equation}
   \rho_O(x;pp) = \langle pp \mid \sum_{i<j}\delta(x-r_{ij})O_{z,+}(ij)
          \mid d,0 \rangle,
\end{equation}
such that
\begin{equation}
    \int_0^\infty\!\! dx \, \rho_O(x)=O\/ \mbox{ contribution}.
\end{equation}
These densities are shown in Fig.~\ref{figrho}, where the $\rho_O(x;pp)$
curves have been rescaled by a {\it single factor\/} $R$, obtained by
matching the maximum of the GT and $pp$ $\Delta \pi$ densities.
As can be seen from Fig.~\ref{figrho}, the GT and $pp$ densities overlap
in the region $x \leq 2$ fm. Of course, at larger $x$-values the
$\rho_O(x;{\rm GT})$ is significantly smaller than the $\rho_O(x;pp)$,
$O=\pi{\rm S}, \Delta \pi, \pi \rho$, because of the increased binding
in the trinucleons. This scaling is to be expected, since it is a
consequence of the ``scaling'' behavior more generally observed for
the calculated $T,S=0,1$ and $T,S=1,0$ pair distribution functions in
nuclei~\cite{For96}; see Figs.~\ref{figrho10} and~\ref{figrho01}. 
Finally, we show in Fig.~\ref{figab} the $\rho_{\Delta \pi}(x)$ densities
obtained with the AV18 and CD-Bonn Hamiltonians for the GT and $pp$
matrix elements. In this case, both the $T=0$ and $T=1$ contributions
are included in the GT densities---namely, they have been calculated
by removing the isospin projector in Eq.~(\ref{p00}). Note that the
$pp$ densities have been rescaled by a factor $R \simeq 39.0$ obtained
by matching the maximum of the AV18 GT and $pp$ densities. However,
this rescaling also makes the CD-Bonn GT and $pp$ densities very close
(see Fig.~\ref{figab}), demonstrating that the $R$ factor has only a very
weak model dependence.

The discussion above shows that two-body contributions to the $pp$
capture are essentially independent of the specific dynamical model
adopted as long as the latter is constrained to reproduce the
experimental value of the GT matrix element.

\section{CONCLUSIONS}
We have calculated the axial matrix element for proton-proton weak
capture using five modern high-precision nucleon-nucleon potentials.
All these models give excellent fits to elastic $N\!N$ scattering data
with a $\chi^2/$datum near 1 and reproduce measured deuteron properties
very well. We have paid particular attention to details of the
electromagnetic interaction and the proper treatment of the low-energy
$pp$ scattering solutions.
As noted before~\cite{Kam94}, the most important correction to the
standard Coulomb interaction between protons is the vacuum polarization,
which reduces the cross section by about 1\%. We have shown that other
fine details of the electromagnetic interaction increase the cross
section by about 0.1\%. This is in part compensated by the correct
relativistic treatment of the deuteron wave number $\gamma$, which
gives a net 0.03\% reduction in the cross section. Including just the
axial one-body operator, the five models differ by only 0.3\% in the
calculated cross section.

The biggest remaining uncertainty is in the contribution of axial two-body
currents, which can increase the cross section by about 1--1.5\%.
Three concerns were expressed at the recent workshop on solar fusion 
rates~\cite{Ade98} regarding the use of the known tritium $\beta$-decay
rate to predict the axial two-body current contribution to the $pp$
fusion reaction: 
(1) the model dependence of the one-body contribution to the
GT matrix element and the resulting uncertainty in the extracted
two-body current contribution to that matrix element;
(2) two-body currents coupling $T,T_z=1,0$ pairs to $T,T_z=1,1$
pairs, which can contribute to the tritium GT matrix element but not to
the $pp$ capture; and
(3) isobar and contact terms could give different contributions
to the GT and $pp$-capture matrix elements, and thus knowledge of their sum 
in the GT may not be sufficient to predict their sum in the capture 
matrix element.

Our detailed calculations show that these concerns do not influence 
the prediction of the $pp$-capture rate. In particular,
(1) the model dependence in the one-body contribution to the
GT matrix element comes mostly from that in the $D$-state probabilities.
Because of the smaller $D$ state predicted by the CD-Bonn potential
(Table~\ref{tabpro}), the corresponding prediction for this contribution
is larger by about 1\% (Table~\ref{tabgt}). 
However, the prediction obtained with this potential for the $pp$-capture 
rate via one-body currents is also larger by about 0.3\%
(Table~\ref{tablamvnn}) because of the smaller $D$ state in the deuteron
(Table~\ref{tabdeutpam}). The axial two-body currents are necessarily
weaker in the CD-Bonn model because they strongly couple the $S$ and
$D$ states. In fact, the sum of one- and two-body current contributions
is much less model dependent than either as can be seen from
Tables~\ref{tabgt} and~\ref{tablpp}.
(2) The axial two-body currents do not couple the $T,T_z=1,0$ 
pairs to the $T,T_z=1,1$ pairs in any significant way, as the discussion
in the preceding section makes clear.
(3) The two-body currents are large at small interparticle
distances where nuclear forces dominate over binding energies. In this
region the pair wave functions in different nuclei are similar in shape
and differ only by a scale factor. This is the basis of the Bethe-Levinger
conjecture~\cite{Lev60}, which can be used to relate processes such as
pion and photon absorption, involving nucleon pairs, in different
nuclei~\cite{For96}. Thus the ratios of GT and $pp$-capture matrix
elements of different two-body current terms are nearly the same as can
be seen from Fig.~\ref{figrho}. Therefore, knowledge of their sum in
the GT matrix element is sufficient to predict their sum in the
$pp$-capture matrix element.

Finally, as we have already mentioned, the GT matrix element is slightly
overpredicted [$\sqrt{3}\times 0.964$ versus the experimental value
$\sqrt{3} \times (0.957 \pm 0.003)$]. Reducing the quark-model prediction
for the $N$ to $\Delta$ axial coupling in the $\Delta\pi$ and $\Delta\rho$
currents by 20\% brings theory and experiment into agreement.
The resulting CD-Bonn and AV18 values for the square of the $pp$ overlap
integral at zero energy are then found to be 7.045 and 7.059, respectively.
Predictions for this quantity with other modern interactions are expected
to fall in this range. Thus, the model dependence and theoretical
uncertainty appear to be at the level of a few parts in a thousand,
much smaller than the estimate given in Ref.~\cite{Ade98}.

\acknowledgments
Several of the authors visited the National Institute for Nuclear Theory
(INT) at the University of Washington in Seattle during the course of
this work and benefitted from discussions with the participants of the
workshop on solar fusion rates and the program on numerical methods for
strongly interacting quantum systems. We would like to thank INT for
the kind hospitality.
The work of J.C.\ and R.S.\ is supported by the U.S.\ Department of
Energy;
that of V.G.J.S.\ and R.B.W.\ is supported by the U.S.\ Department of
Energy, Nuclear Physics Division, under Contract No.\ W-31-109-ENG-38;
that of W.G., H.K., and A.N.\ is supported by the Deutsche
Forschungsgemeinschaft and the Research Contract No.\ 41324878 (COSY-044)
of the Forschungszentrum J\"ulich;
that of R.M.\ and V.R.P.\ is supported by the U.S.\ National Science
Foundation via Grant No.\ PHY96-03097 and Grant No.\ PHY94-21309,
respectively;
finally, the work of A.K., S.R., and M.V.\ is supported by the Italian
Istituto Nazionale di Fisica Nucleare.
The calculations were made possible by grants of computer time from
the National Energy Research Supercomputer Center and
Hoechstleitungsrechenzentrum J\"ulich.

\end{multicols}

\appendix
\section{THE AXIAL TWO-BODY CURRENT OPERATORS}
For completeness, we list here the momentum-space expressions for the
axial two-body currents used in the present work.

\noindent (1) Axial $\pi$-exchange $\Delta$-excitation current:
\begin{equation}
  {\bf A}^{(2)}_{a,ij} ({\bf q};\Delta \pi)  = 
        - {16 \over 25} \, g_A  \, {f_{\pi N\!N}^2 \over
           m_\pi^2 (m_\Delta - m)} \, { {\bbox \sigma}_j \cdot {\bf k}_j
           \over m_\pi^2 + k_j^2} \, f_{\pi}^2 (k_j) 
         \left[ 4\, \tau_{j,a} \, {\bf k}_j
     -({\bbox \tau}_i \times {\bbox \tau}_j)_a \,
     {\bbox \sigma}_i \times {\bf k}_j  \right] + i \rightleftharpoons j.
\end{equation}

\noindent (2) Axial $\rho$-exchange $\Delta$-excitation current:
\begin{equation}
  {\bf A}^{(2)}_{a,ij} ({\bf q};\Delta \rho)  = 
         {4 \over 25} \, g_A \, {g_\rho^2 (1 + \kappa_\rho)^2
         \over m^2 (m_\Delta - m)} \, {f_{\rho}^2(k_j) \over
          m_\rho^2 + k_j^2} \left\{ 4 \, \tau_{j,a} \,
         ( {\bbox \sigma}_j \times {\bf k}_j) \times {\bf k}_j
     -( {\bbox \tau}_i \times {\bbox \tau}_j)_a \,
      {\bbox \sigma}_i \times  \left [ ( {\bbox \sigma}_j \times {\bf k}_j) 
      \times {\bf k}_j \right ] \right\} + i \rightleftharpoons j.
\end{equation}

\noindent (3) Axial $\pi$-exchange (pair) current:
\begin{equation}
  {\bf A}^{(2)}_{a,ij}({\bf q}; \pi {\rm S}) = 
        {g_A \over 2 m} \, {f_{\pi N\!N}^2 \over m_\pi^2}
     \, { \bbox{\sigma}_j \cdot {\bf k}_j \over m_\pi^2 + k_j^2}
        f_\pi^2(k_j) \left\{ (\bbox{\tau}_i \times \bbox{\tau}_j)_a
        \bbox{\sigma}_i \times {\bf k}_j
        -i\tau_{j,a} \left [ {\bf q}+i
        \bbox{\sigma}_i \times ( {\bf p}_i + {\bf p}^\prime_i ) \right ]
        \right\} + i \rightleftharpoons j.
\end{equation}

\noindent (4) Axial $\rho$-exchange (pair) current:
\begin{eqnarray}
   {\bf A}^{(2)}_{a,ij}({\bf q}; \rho {\rm S}) = &&
    -  g_A {g_\rho^2 (1 + \kappa_\rho)^2 \over 8m^3}
       {f_\rho^2 (k_j) \over m_\rho^2 + k_j^2} \left( \tau_{j,a}
       \left\{ (\bbox{\sigma}_j \times {\bf k}_j) \times {\bf k}_j
    -  i \left[ \bbox{\sigma}_i \times (\bbox{\sigma}_j \times
       {\bf k}_j ) \right] \times ( {\bf p}_i + {\bf p}^\prime_i) \right\}
                                               \right.  \nonumber\\
   &&  \left. + (\bbox{\tau}_i \times \bbox{\tau}_j)_a \left\{ {\bf q} 
       \bbox{\sigma}_i \cdot ( \bbox{\sigma}_j \times {\bf k}_j)
    +  i (\bbox{\sigma}_j \times {\bf k}_j) \times
       ({\bf p}_i + {\bf p}^\prime_i) 
       - \left[ \bbox{\sigma}_i \times (\bbox{\sigma}_j \times {\bf k}_j)
       \right] \times {\bf k}_j \right\} \right) + i \rightleftharpoons j.
\end{eqnarray}

\noindent (5) Axial $\pi\rho$ current:
\begin{equation}
   {\bf A}^{(2)}_{a,ij}({\bf q}; \pi \rho) =
   -  g_A  {g_\rho^2 \over m} \, { \bbox{\sigma}_j \cdot {\bf k}_j \over 
       (m_\rho^2 + k_i^2) (m_\pi^2 + k_j^2)}
       f_\rho (k_i) f_\pi (k_j) (\bbox{\tau}_i \times \bbox{\tau}_j)_a
      \left[ (1 + \kappa_\rho) \bbox{\sigma}_i \times {\bf k}_i  
      - i ({\bf p}_i + \ {\bf p}^\prime_i ) \right]
      + i \rightleftharpoons j.
\end{equation}
Here ${\bf q}$ is the total momentum transfer,
${\bf q}={\bf k}_i+{\bf k}_j$,
${\bf k}_{i(j)}$ is the momentum transfer to nucleon $i$ ($j$),
${\bf p}_i$ and ${\bf p}^\prime_i$ are the initial and final momenta
of nucleon $i$, and $f_{ \pi (\rho)} (k)$=pion ($\rho$-meson)-nucleon 
monopole vertex form factor. The quark model has been used to relate
the $\pi N \Delta$, $\rho N \Delta$ and axial $N \Delta$ couplings to,
respectively, the $\pi N\!N$, $\rho N\!N$, and $g_A$ couplings.
The expression for $\pi S$ represents the conventional pair current
operator given in the literature. It is obtained with pseudoscalar
pion-nucleon coupling. With pseudovector coupling the pion momentum
${\bf k}_j$ in the first term in brackets would be replaced by the
external momentum ${\bf q}$ and an additional term
(${\bf p}_i+{\bf p}^\prime_i$) would appear with the isospin structure
$(\bbox{\tau}_i \times \bbox{\tau}_j)_a$. Furthermore, the $\rho S$ 
operator includes only those terms which are proportional to
$(1+\kappa_\rho)^2$. Finally, $m_\pi$, $m_\rho$, $m$, and $m_\Delta$
are, respectively, the pion, $\rho$-meson, nucleon, and $\Delta$ masses.

\begin{multicols}{2}

\end{multicols}

\pagebreak

\mediumtext
\begin{table}
\caption{Triplet $S$-wave low-energy scattering parameters
         and deuteron properties.}
\begin{tabular}{lcccccc}
        &  AV18  & CD-Bonn& Nijm-I & Nijm-II& Reid93 & Empirical \\
\tableline
$a_t$ (fm)
        & 5.419  & 5.419  & 5.418  & 5.420  & 5.422  & 5.419(7)$^a$ \\
$r_t=\rho(0,0)$ (fm)
        & 1.753  & 1.752  & 1.751  & 1.753  & 1.755  & 1.754(8)$^a$ \\
$\rho_d=\rho(-B_d,-B_d)$ (fm)
        & 1.767  & 1.764  & 1.762  & 1.764  & 1.769  & 1.765(5)$^b$ \\
$A_S$ (fm$^{-1/2}$)
        & 0.8850 & 0.8845 & 0.8841 & 0.8845 & 0.8853 & 0.8845(9)$^b$ \\
$\eta$  & 0.0250 & 0.0255 & 0.0253 & 0.0252 & 0.0251 & 0.0256(4)$^c$ \\
$r_d$ (fm)
        & 1.967  & 1.966  & 1.967  & 1.968  & 1.969  & 1.97535(85)$^d$ \\
$Q_d$ (fm$^2$)
        & 0.270  & 0.270  & 0.272  & 0.271  & 0.270  & 0.2859(3)$^e$ \\
$P_D$ (\%)
        & 5.76   & 4.83   & 5.66   & 5.64   & 5.70   & ---
\end{tabular}
$^a$Reference~\cite{Hou71}.\\
$^b$References~\cite{Eri83,Swa95}.\\
$^c$Reference~\cite{Rod90}.\\
$^d$Reference~\cite{Hub98}.\\
$^e$References~\cite{Bis79,Eri83}.
\label{tabdeutpam}
\end{table}

\narrowtext
\begin{table}
\caption{Square of the overlap integral $\Lambda(E_{\rm lab})$ at various
         laboratory energies for the five $N\!N$ potential models.
         The zero-energy results are obtained by extrapolating the
         preceding results.}
\begin{tabular}{lcccccc}
  $N\!N$ model &  Ref.  &  5 keV &  4 keV &  3 keV &  2 keV &  0 keV \\
\tableline
 AV18    & \cite{Wir95} & 7.002 & 6.995 & 6.987 & 6.980 & 6.965 \\
 CD-Bonn & \cite{Mac96} & 7.022 & 7.014 & 7.007 & 6.999 & 6.985 \\
 Nijm~I  & \cite{Sto94} & 7.002 & 6.994 & 6.987 & 6.979 & 6.965 \\
 Nijm~II & \cite{Sto94} & 7.008 & 7.000 & 6.993 & 6.986 & 6.971 \\
 Reid93  & \cite{Sto94} & 7.011 & 7.003 & 6.996 & 6.989 & 6.974
\end{tabular}
\label{tablamvnn}
\end{table}

\narrowtext
\begin{table}
\caption{Square of the overlap integral $\Lambda(E_{\rm lab})$ at various
         laboratory energies for four different truncations of the
         electromagnetic interaction (all for point-particle protons).
         The nuclear interaction is the AV18 potential~\protect\cite{Wir95}.
         The result for the full interaction with finite-size
         contributions is included for comparison.}
\begin{tabular}{lccccc}
 $V_{\rm EM}(pp)$           & 5 keV & 4 keV & 3 keV & 2 keV & 0 keV \\
\tableline
 $V_{C1}$                   & 7.060 & 7.051 & 7.043 & 7.035 & 7.019 \\
 $V_{C1}+V_{C2}$            & 7.063 & 7.055 & 7.047 & 7.039 & 7.023 \\
 $V_{C1}+V_{\rm VP}$        & 6.993 & 6.985 & 6.978 & 6.971 & 6.956 \\
 $V_{C1}+V_{C2}+V_{\rm VP}$ & 6.996 & 6.989 & 6.981 & 6.974 & 6.960 \\
 Full                       & 7.002 & 6.995 & 6.987 & 6.980 & 6.965
\end{tabular}
\label{tabtrunc}
\end{table}

\narrowtext
\begin{table}
\caption{One-body and two-body $\Delta \pi$ contributions to the
         Gamow-Teller matrix element of tritium $\beta$ decay, obtained
         with various combinations of modern two- and three-nucleon
         interactions in CHH and 42-channel Faddeev calculations,
         the former for the AV18/UIX model only. The one-body results
         obtained from Eq.~(\protect\ref{ppp}) are also quoted, while
         those under the heading ``Total'' give the sum of the
         one-body (first column) and $\Delta \pi$ contributions.}
\begin{tabular}{lcccc}
  Hamiltonian  & One-Body & Eq.~(\protect\ref{ppp}) & $\Delta\pi$ & Total \\
\tableline
 AV18          & 0.924 &    0.925    &   0.0507     & 0.975  \\
 AV18/TM       & 0.925 &    0.925    &   0.0546     & 0.980  \\
 AV18/UIX      & 0.922 &    0.923    &   0.0560     & 0.979  \\
\tableline
 CD-Bonn       & 0.935 &    0.935    &   0.0427     & 0.977  \\
 CD-Bonn/TM    & 0.937 &    0.937    &   0.0435     & 0.980  \\
\tableline
 Nijm~I        & 0.926 &    0.927    &   0.0507     & 0.977  \\
 Nijm~I/TM     & 0.928 &    0.927    &   0.0534     & 0.981  \\
\tableline
 Nijm~II       & 0.926 &    0.927    &   0.0504     & 0.976  \\
 Nijm~II/TM    & 0.927 &    0.927    &   0.0534     & 0.981  \\
\tableline
 Reid93        & 0.925 &    0.926    &   0.0514     & 0.977  \\
 Reid93/TM     & 0.926 &    0.926    &   0.0549     & 0.981     
\end{tabular}
\label{tabgt}
\end{table}

\narrowtext
\begin{table}
\caption{The $S$-, $S^\prime$-, $P$-, and $D$-state percent probabilities
         in $^3$H wave functions. The results for the AV18/UIX model
         are from Ref.~\protect\cite{Kie95}.}
\begin{tabular}{lcccc}
  Hamiltonian       &  $S$   & $S^\prime$ & $P$     & $D$     \\
\tableline
 AV18               & 90.10  & 1.33       & 0.066   & 8.51    \\
 AV18/TM            & 89.96  & 1.09       & 0.155   & 8.80    \\
 AV18/UIX           & 89.51  & 1.05       & 0.130   & 9.31    \\
\tableline
 CD-Bonn            & 91.62  & 1.34       & 0.046   & 6.99    \\
 CD-Bonn/TM         & 91.74  & 1.21       & 0.102   & 6.95    \\
\tableline
 Nijm~I             & 90.29  & 1.27       & 0.066   & 8.37    \\
 Nijm~I/TM          & 90.25  & 1.08       & 0.148   & 8.53    \\
\tableline
 Nijm~II            & 90.31  & 1.27       & 0.065   & 8.35    \\
 Nijm~II/TM         & 90.22  & 1.07       & 0.161   & 8.54    \\
\tableline
 Reid93             & 90.21  & 1.28       & 0.067   & 8.44    \\
 Reid93/TM          & 90.09  & 1.07       & 0.162   & 8.68     
\end{tabular}
\label{tabpro}
\end{table}

\narrowtext
\begin{table}
\caption{Contributions to the Gamow-Teller matrix element of tritium
         $\beta$ decay, obtained with the CHH AV18/UIX trinucleon wave
         functions. The cutoff masses $\Lambda_\pi=4.8$ fm$^{-1}$ and
         $\Lambda_\rho=6.8$ fm$^{-1}$ are used in the axial two-body
         operators. The cumulative result is 0.9636. The two-body
         results obtained by retaining only the contributions of the
         $T=1$ pairs in tritium are also given (column labelled $T=1$).}
\begin{tabular}{lcc}
               &  Total    &  $T$=1   \\
\tableline
One-body       &   0.9218  &          \\
$\Delta \pi$   &   0.0560  &   0.0291 \\
$\Delta \rho$  & --0.0213  & --0.0111 \\
$\pi \rho$     &   0.0070  &   0.0035 \\
$\pi S$        &   0.0044  &   0.0025 \\
$\rho S$       & --0.0043  & --0.0021 
\end{tabular}
\label{tabgt2}
\end{table}

\mediumtext
\begin{table}
\caption{Square of the overlap integral $\Lambda(E_{\rm lab})$ at various
         laboratory energies for the AV18 and CD-Bonn interactions.
         The zero-energy results are obtained by linear extrapolation of
         those at $E_{\rm lab}=3$ and 5 keV. The cutoff masses
         $\Lambda_\pi=4.8$ fm$^{-1}$ and $\Lambda_\rho=6.8$ fm$^{-1}$
         are used in the axial two-body operators. The two-body
         contributions are added successively in the given order.}
\begin{tabular}{lcccccccc}
       & \multicolumn{2}{c}{5 keV} & \multicolumn{2}{c}{4 keV}
       & \multicolumn{2}{c}{3 keV} & \multicolumn{2}{c}{0 keV} \\
       & AV18 & CD-Bonn & AV18 & CD-Bonn & AV18 & CD-Bonn & AV18 & CD-Bonn \\
\tableline
One-body      & 7.002 & 7.022 & 6.995 & 7.014 & 6.987 & 7.007 & 6.965 & 6.985 \\
$+\pi$S       & 7.015 & 7.024 & 7.007 & 7.016 & 6.999 & 7.009 & 6.977 & 6.987 \\
$+\rho$S      & 7.005 & 7.018 & 6.997 & 7.010 & 6.990 & 7.003 & 6.967 & 6.981 \\ 
$+\Delta\pi$  & 7.138 & 7.126 & 7.130 & 7.118 & 7.122 & 7.111 & 7.099 & 7.089 \\
$+\Delta\rho$ & 7.090 & 7.092 & 7.083 & 7.084 & 7.075 & 7.077 & 7.052 & 7.055 \\
$+\pi\rho$    & 7.114 & 7.097 & 7.107 & 7.089 & 7.099 & 7.082 & 7.076 & 7.060
\end{tabular}
\label{tablpp}
\end{table}

\narrowtext
\begin{figure}
\caption{Deuteron wave functions: large curves, $u(r)$;
         small curves, $w(r)$. The solid, dashed, dash-dotted, dotted,
         and long-dashed curves are generated from the
         CD-Bonn, Nijm-I, Nijm-II, Reid93, and AV18 potentials,
         respectively.}
\label{figdeut}
\end{figure}

\narrowtext
\begin{figure}
\caption{Gamow-Teller (solid lines) and $pp$ (dashed lines)
         two-body densities. Note that all $pp$ curves have been
         rescaled by a single factor, as explained in the text.}
\label{figrho}
\end{figure}

\narrowtext
\begin{figure}
\caption{The $T,S=1,0$ pair distribution functions for various nuclei;
         see Ref.~\protect\cite{For96}. Note that the curves have been
         renormalized to the peak height of the $^{16}$O density.}
\label{figrho10}
\end{figure}

\narrowtext
\begin{figure}
\caption{The $T,S=0,1$ $M_S=0,\pm1$ pair distribution functions for
         given angles $\theta$ between the spin-quantization axis and the
         relative position vector of the two nucleons and for various
         nuclei; see Ref.~\protect\cite{For96}. Note that the curves
         have been renormalized to the peak height of the deuteron
         $M_S=\pm1$ $\theta=0$ density.}
\label{figrho01}
\end{figure}

\narrowtext
\begin{figure}
\caption{Gamow-Teller (solid lines) and $pp$ (dashed lines) $\Delta\pi$
         densities obtained with the AV18 and CD-Bonn Hamiltonians.
         Note that the Gamow-Teller densities include both the $T=0$ and
         $T=1$ contributions---namely, they have been calculated by
         removing the isospin projector $P^\tau_0(ij)$ in
         Eq.~(\protect\ref{p00}). The $pp$ densities have been rescaled
         by a single factor $R\simeq 39.0$, obtained by matching the
         maximum of the AV18 Gamow-Teller and $pp$ densities.}
\label{figab}
\end{figure}

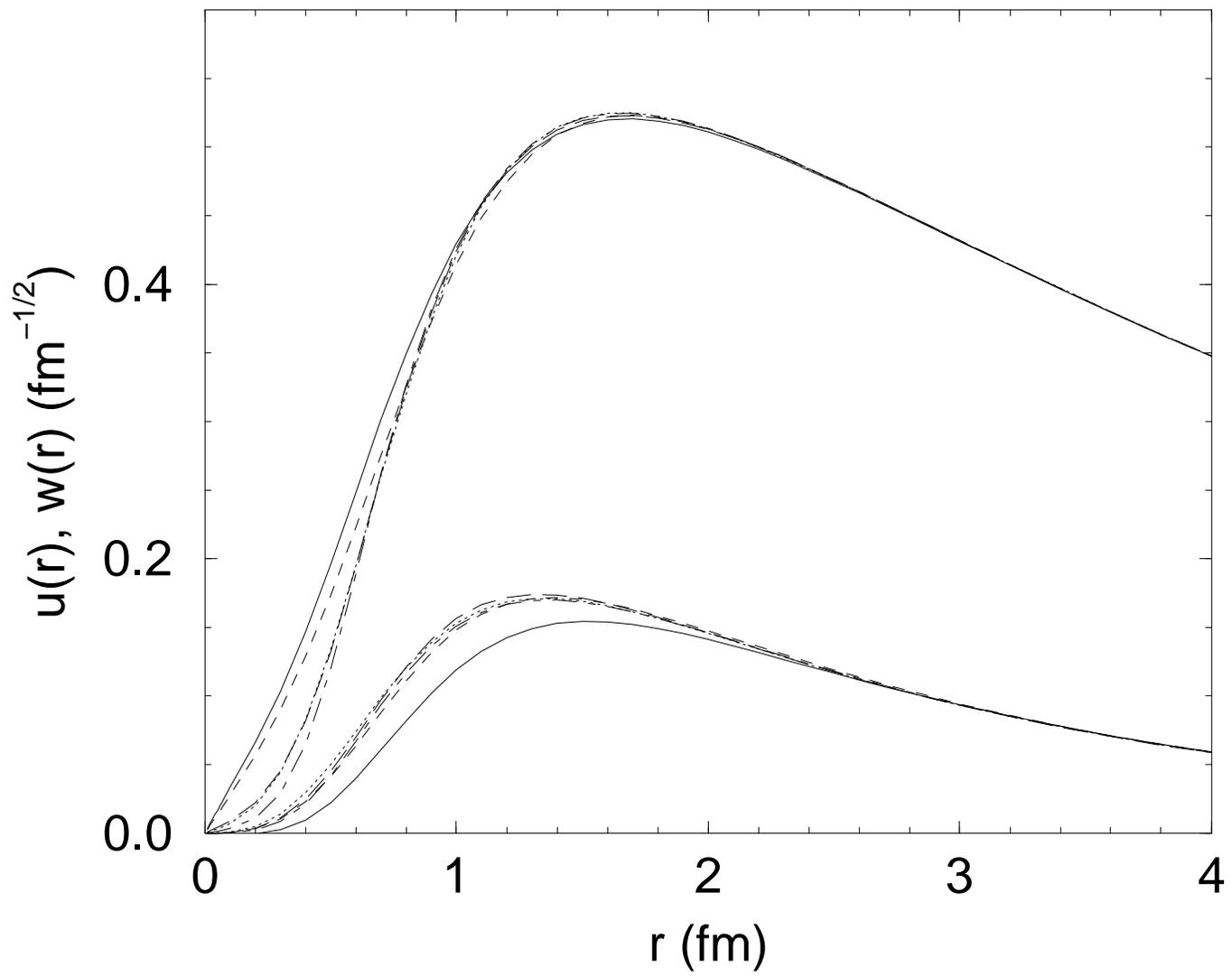
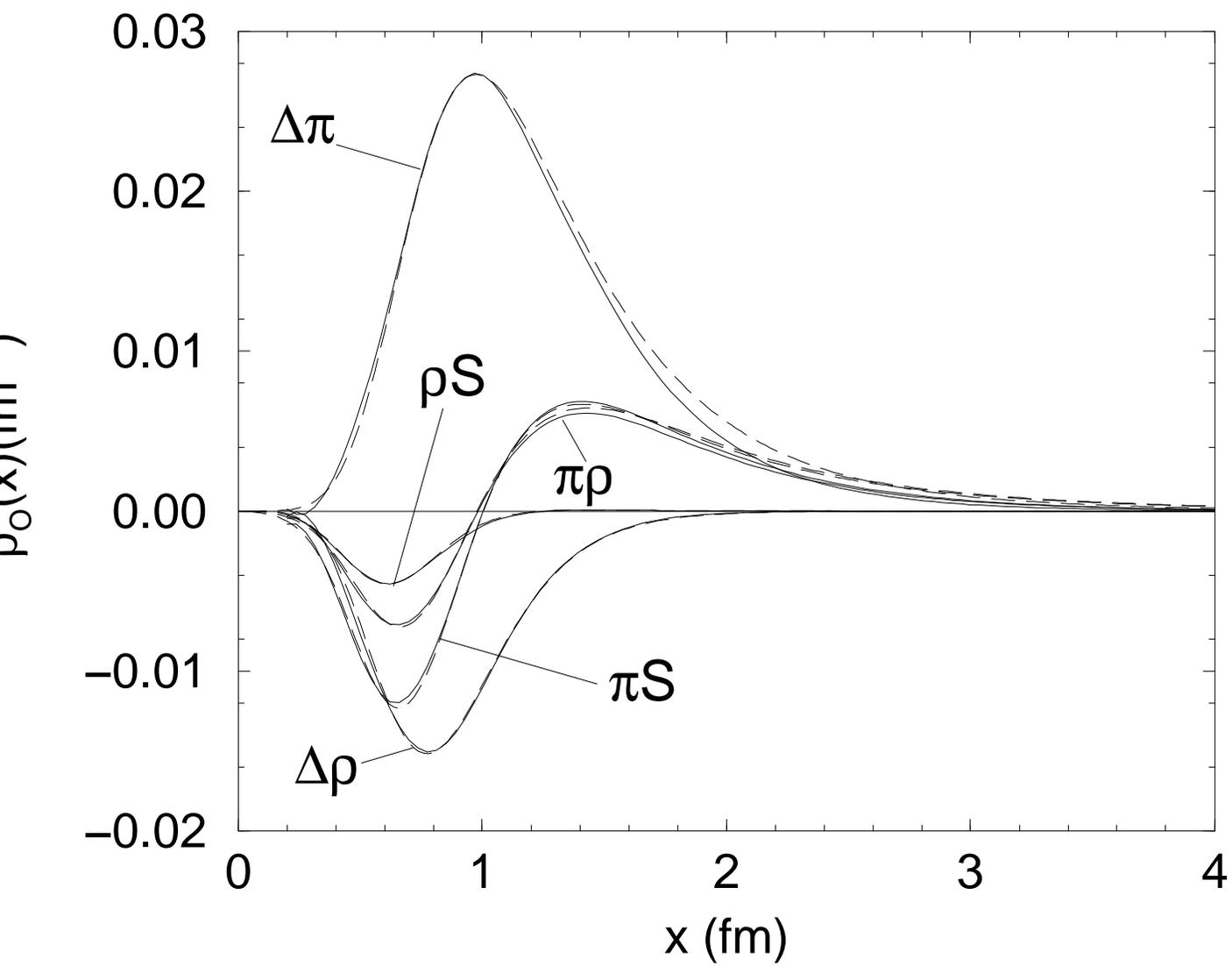
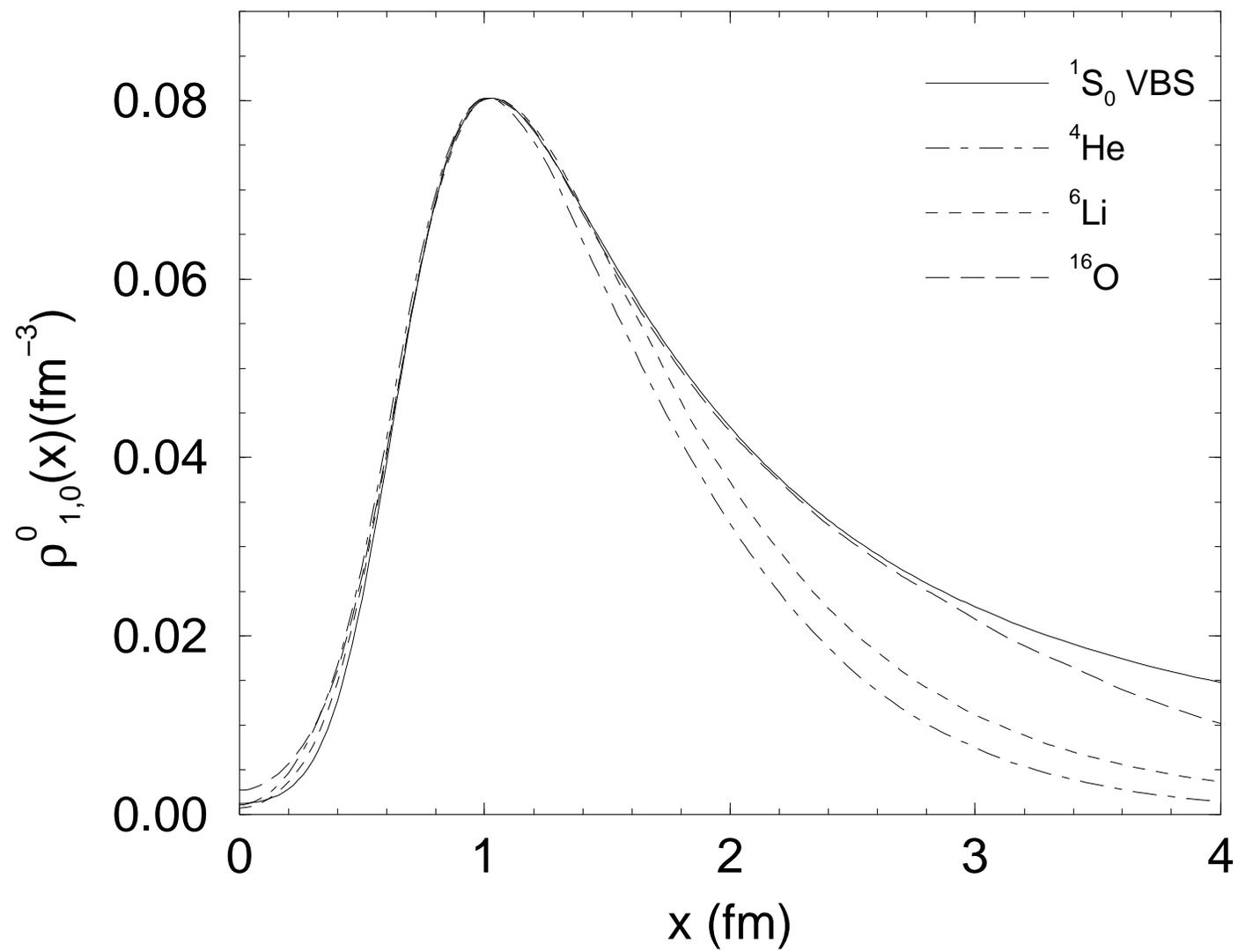
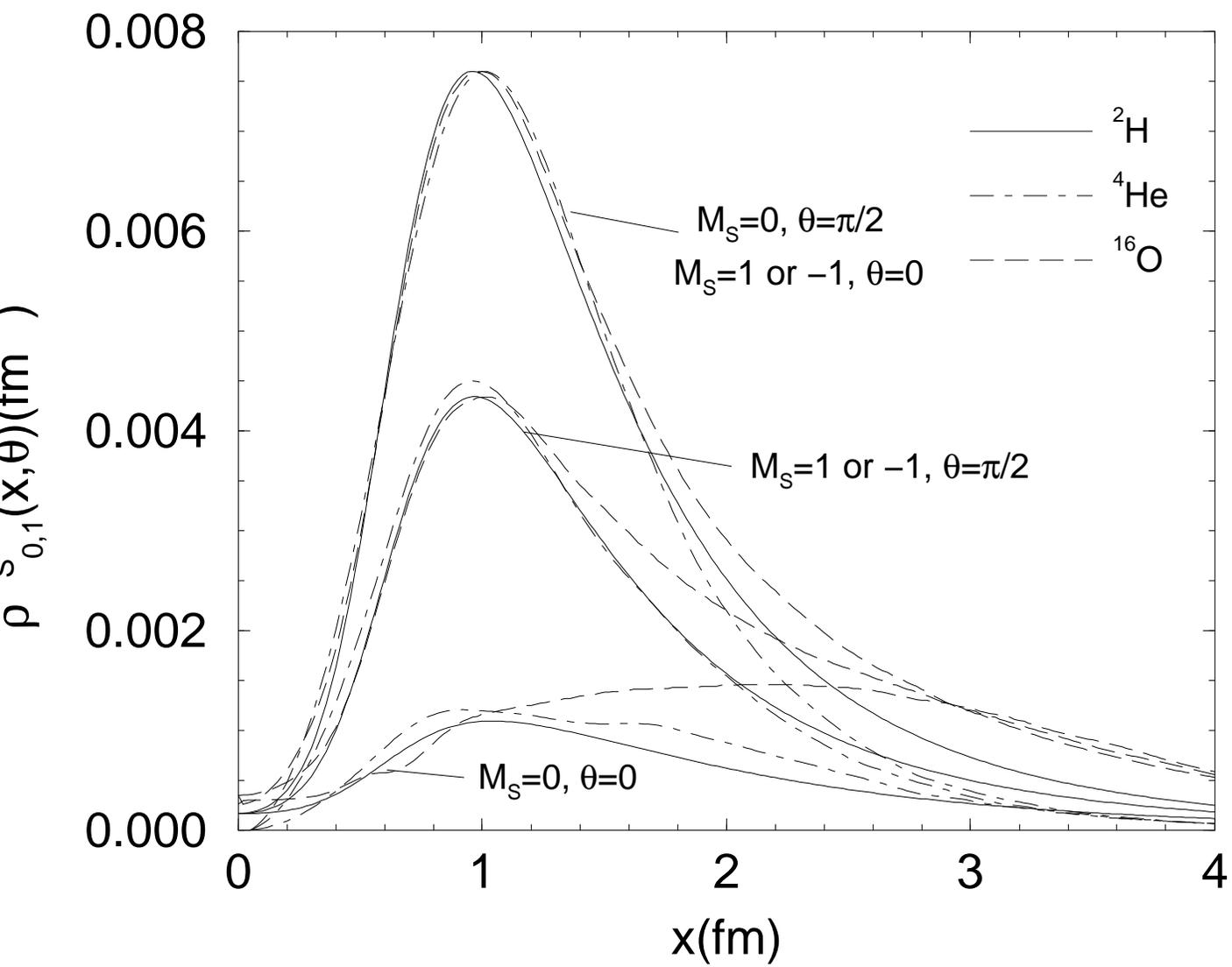
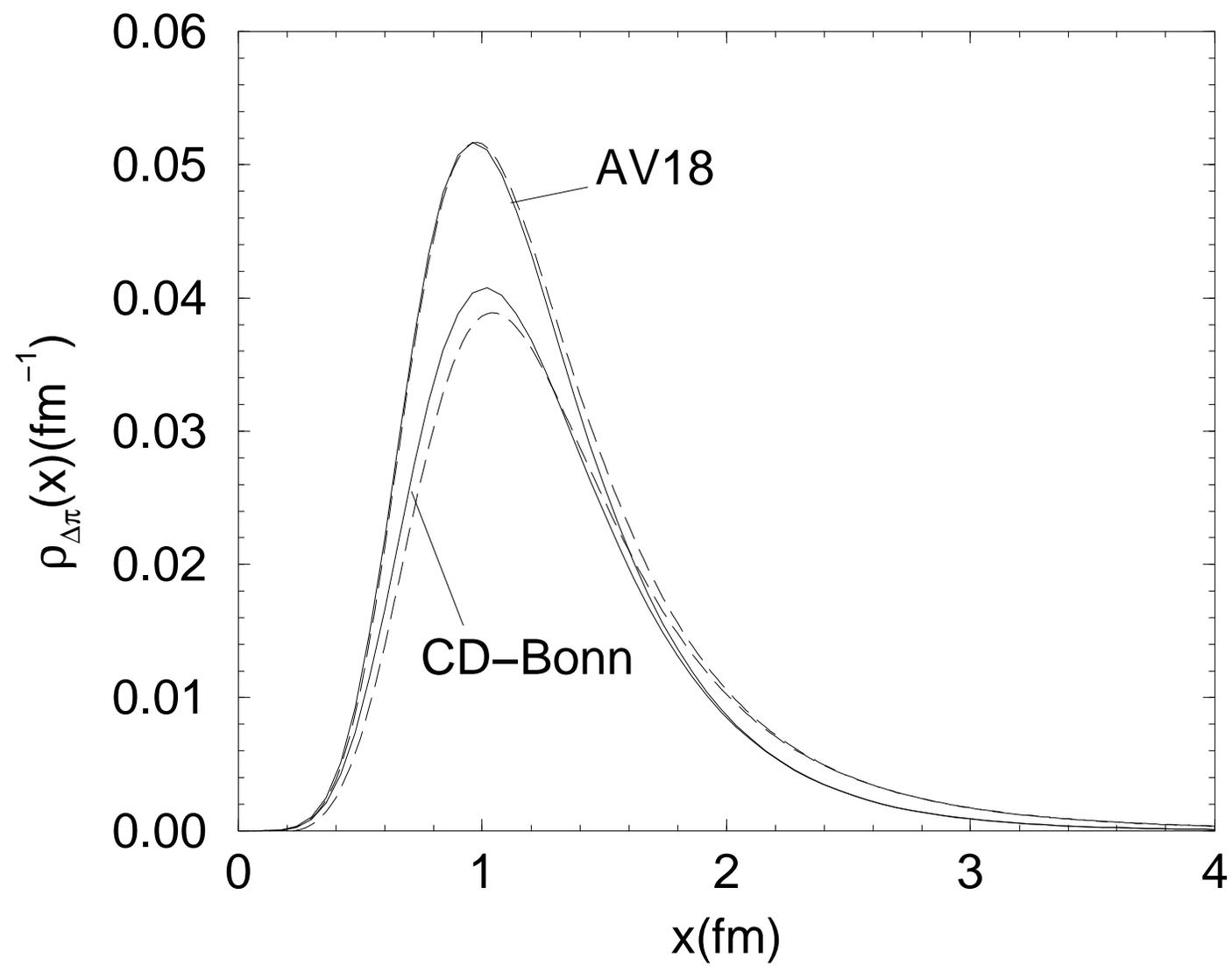
\end{document}